\documentclass[12pt,preprint,dvips,deluxetable]{aastex}

\def \co{$^{12}$CO}
\def \msun{$\mathrm{M}_\odot$}

\def \kms{km~s$^{-1}$}
\def \1pap{Paper I}
\def \pap2{Paper II}

\makeatletter
\newcommand\cutinheadb[1]{%
 \noalign{\vskip .8ex}%
 \@ptabularcr
 \noalign{\vskip -4ex}%
 \multicolumn{\pt@ncol}{c}{#1}%
 \@ptabularcr
 \noalign{\vskip .8ex}%
 \hline
 \@ptabularcr
 \noalign{\vskip -1.5ex}%
}%
\def\cutinheadb@ppt#1{%
 \noalign{\vskip .8ex}%
 \@ptabularcr
 \noalign{\vskip -1.5ex}
 \multicolumn{\pt@ncol}{c}{#1}%
 \@ptabularcr
 \noalign{\vskip .8ex}%
 \hline
 \@ptabularcr
 \noalign{\vskip -1.5ex}%
}%
\makeatother

\makeatletter
\newcommand\cutinheadc[1]{%
 \noalign{\vskip .8ex}%
 \@ptabularcr
 \noalign{\vskip -4ex}%
 \multicolumn{\pt@ncol}{c}{#1}%
 \@ptabularcr
 \noalign{\vskip .8ex}%
 \hline
 \@ptabularcr
 \noalign{\vskip -1.5ex}%
}%
\def\cutinheadc@ppt#1{%
 \noalign{\vskip .8ex}%
 \@ptabularcr
 \noalign{\vskip -1.5ex}
 \multicolumn{\pt@ncol}{c}{#1}%
 \@ptabularcr
 \noalign{\vskip .8ex}%
 \hline
 \@ptabularcr
 \noalign{\vskip -1.5ex}%
}%
\makeatother

\author{Henry~A.~Kobulnicky\altaffilmark{1},
Rachel A. Smullen\altaffilmark{1},
Daniel~C.~Kiminki\altaffilmark{2},
Jessie C. Runnoe\altaffilmark{1},
Earl S. Wood\altaffilmark{1},
Garrett Long\altaffilmark{1},
Michael J. Alexander\altaffilmark{1},
Michael J. Lundquist\altaffilmark{1},
Carlos Vargas-Alvarez\altaffilmark{1}}

\altaffiltext{1}{Dept. of Physics \& Astronomy, University 
of Wyoming, Laramie, WY 82070 \\ chipk@uwyo.edu}
\altaffiltext{2}{Dept. of Astronomy, University of Arizona,
Tucson, AZ 85721} 

\begin{document}
\slugcomment{To be published in The Astrophysical Journal}

\title{A Fresh Catch of Massive Binaries in 
the Cygnus OB2 Association}

\begin{abstract}

Massive binary stars may constitute a substantial fraction of
progenitors to supernovae and $\gamma$-ray bursts, and the
distribution of their orbital characteristics holds clues to the
formation process of massive stars.  As a contribution to securing
statistics on OB-type binaries, we report the discovery and orbital
parameters for five new systems as part of the Cygnus~OB2 Radial
Velocity Survey.  Four of the new systems (MT070, MT174, MT267, and
MT734$\equiv$VI Cygni \#11) are single-lined spectroscopic binaries
while one (MT103) is a double-lined system (B1V+B2V).  MT070 is
noteworthy as the longest period system yet measured in Cyg~OB2, with
$P$=6.2~yr.  The other four systems have periods ranging between 4 and
73~days.  MT174 is noteworthy for having a probable mass ratio
$q<0.1$, making it a candidate progenitor to a low-mass X-ray binary.
These measurements bring the total number of massive binaries in
Cyg~OB2 to 25, the most currently known in any single cluster or
association.

\end{abstract}

\keywords{Techniques: radial velocities --- (Stars:) binaries: general --- (Stars:) binaries: spectroscopic --- (Stars:) binaries:
 (\textit{including multiple}) close --- Stars: early-type --- Stars:
 kinematics and dynamics}

\section{Introduction}

Massive stars produce some of nature's most energetic phenomena as
they end their lives in spectacular supernova or $\gamma$-ray burst
explosions \citep{burrows95,woosleylangerweaver,woosley2006}.  While
the post-main-sequence evolutionary path of a single massive star is
still not settled (e.g., the sequence through red/blue supergiant,
Wolf-Rayet, Luminous Blue Variable phases for stars of various
masses; \citealt[][]{meynet2000}), the possible evolutionary paths of
close massive binary systems are even less certain.  Nevertheless,
close massive binaries appear common
\citep{garmany80,sanaevans2011,Kiminki2012a}, and secondary stars in
such systems have been implicated in removing hydrogen envelopes from
massive primaries prior to core collapse, resulting in a substantial
fraction (30--40\%) of Type Ib/c supernovae and $\gamma$-ray bursts
\citep{izzard2004, kf07, eldridge2008}. Close massive binaries may
also produce the observed population of high-velocity ``runaway''
stars either through ejection of one component after the supernova
explosion or during multi-body gravitational interactions
\citep{giesbolton, blaauw}.

Papers I--IV in this series (\citealt{Kiminki07},
\citealt{Kiminki08}, \citealt{Kiminki09}, \citealt{Kiminki2012a})
describe prior results of the Cygnus~OB2 Radial Velocity Survey
intended to measure the massive binary characteristics (i.e., binary
fraction, distribution of periods, mass ratios, eccentricities) for a
large number (114) massive stars in a single cluster/association
having a common formation environment.  In particular, Paper IV
(Table~6) summarizes orbital elements for the 20 massive binaries
with measured parameters.  Paper V \citep{Kiminki2012b} uses these
previously published data and the new data presented herein (a total
of 25  measured systems) to infer the intrinsic distributions of
massive binaries, concluding that the fraction of massive stars
(defined as B3 and earlier; i.e., supernova progenitors\footnote{For
a typical initial mass function \citep{salpeter55}, B stars outnumber
O stars 3:1 as supernova progenitors.}) having companions may be as
high as 90\%, and that 45\% of these are likely to interact at some
point.   Paper V reports an excess of short-period 4--7 day systems
relative to 7--14 day systems and finds that the distribution of mass
ratios is approximately flat over the range 0.1$<q<$1.0. Besides
providing the fundamental data for modeling the frequencies of
energetic phenomena, these types of statistics help place constraints
on theoretical frameworks for the formation of massive stars which
remain under debate \citep{Krumholzetal2010, smithetal2009}.

\begin{figure}
\epsscale{1.0}
\centering
\plotone{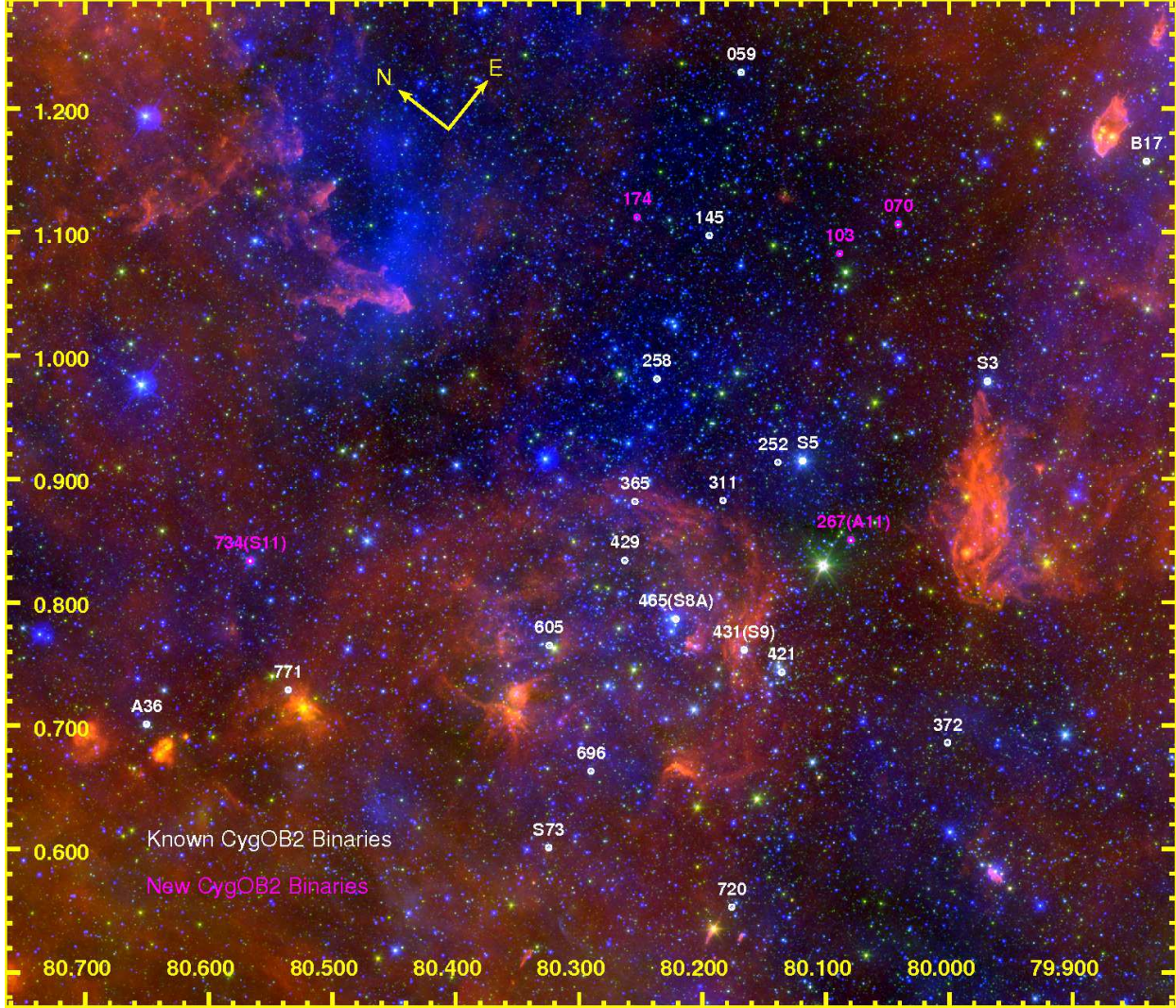}
\caption{Three-color image of the Cygnus~OB2 region with the Palomar Sky Survey
R image in blue, the \emph{Spitzer} 4.5 $\mu$m in green, and \emph{Spitzer} 8.0 $\mu$m in 
red.  White labels denote previously known binary systems while magenta labels
highlight the newly discovered systems reported herein. Generally, numeration follows the
system of \citet{MT91}, with ``S'' additionally indicating the
numeration of \citet{Schulte58} and ``A'' or ``B'' for that of \citet{comeron02}.
The early-B SB2 system A45 lies just off the field of view to the upper right of B17.   
\label{color}}
\end{figure}

In this sixth paper of the series we present orbital solutions for
five additional massive spectroscopic binaries in Cyg~OB2---one
double-lined (SB2) and four single-lined (SB1) systems---using
spectroscopic data obtained primarily during 2010--2011, but utilizing
some data as early as 1999.  Using the nomenclature of \citet{MT91},
these systems are MT103 (SB2), MT070, MT174, MT267, and MT734 (SB1s).
Figure~\ref{color} displays a three-color representation of the
Cyg~OB2 vicinity, where blue/green/red depict the Palomar Sky Survey
R, the $Spitzer$ 4.5~$\mu$m, and the $Spitzer$ 8.0~$\mu$m bands,
respectively.  White labels denote previously known binary systems,
while magenta labels highlight the newly discovered systems reported
herein. Numeration follows the system of \citet{MT91}, with ``S''
additionally indicating the numeration of \citet{Schulte58} and ``A''
or ``B'' for that of \citet{comeron02}.  Figure~\ref{color} shows the
complex nature of this region, including ionized gas (diffuse blue
emission tracing H$\alpha$ within the POSS R band),
photo-dissociation regions (PDR) at the edges of molecular
clouds (diffuse red and green tracing broad emission features arising
from excited polycyclic aromatic hydrocarbons, PAHs), and stars (blue
and green point sources).

After first reviewing the Survey strategy and
data, we present the new radial velocity data and describe the
orbital characteristics of these O and early B systems.  We conclude
by summarizing the parameters of the 25 currently known massive
binaries in CygOB2.  All reported velocities are in the heliocentric 
frame unless explicitly indicated otherwise.

\section{Spectroscopic Observations, Reductions, and Radial Velocity 
Measurement}

The Cygnus~OB2 Radial Velocity Survey is an ongoing optical
spectroscopic survey of $\sim$120 massive primary stars (earlier than
B3) drawn from the \citet{MT91} photometric study of Cyg~OB2.  Papers
I--IV provide details regarding the variety of observatories, dates
of observation, and data analysis methods employed dating back to the
start of the survey in 1999.  Observing cadences vary from nightly to
as few as several observations per year for the survey overall. 
Here, we repeat essential observational details and report on new
observations covering the 2010 and 2011 observing seasons at the
Wyoming Infrared Observatory (WIRO) 2.3~m telescope.
Table~\ref{103.tab} provides a log of observation dates and derived
radial velocities for the SB2 MT103 and Table~\ref{rest.tab} lists
observing dates and velocities for the four SB1 systems.

In 2011, the data were obtained using the WIRO Longslit spectrograph
with an e2V 2048$^2$ ccd as the detector.  An 1800~l~mm$^{-1}$
grating in first order yielded a spectral resolution of 1.5~\AA\ near
5800~\AA\ with a 2\arcsec$\times$100\arcsec\ slit.   The spectral
coverage was 5250--6750~\AA. Exposure times ranged from 1200 s to
3600 s in multiples of 600 s depending on target brightness, current
seeing (1.2--3 arcseconds FWHM) and cloud conditions.  Reductions
followed standard longslit techniques, including flat fielding from
dome quartz lamp exposures. Copper-argon arc lamp exposures were
taken after each star exposure to wavelength calibrate the spectra to
an rms of 0.03~\AA\ (1.5~\kms\ at 5800~\AA). Multiple exposures were
combined yielding final S/N ratios typically in excess of 100:1 near
5800~\AA. Final spectra were Doppler corrected to the heliocentric
frame.  Each spectrum was then shifted by a small additional amount
in velocity so that the Na~I~D $\lambda\lambda$5890,5996 lines were
registered with the mean Na I line wavelength across the ensemble of
observations. This zero-point correction to each observation is
needed to account for effects of image wander in the dispersion
direction when the stellar FWHM of the point spread function 
was appreciably less than the slit
width.  Because of these inevitable slit-placement effects on
the resulting wavelength solutions at the level of $\lesssim$10 \kms,
radial velocity standards were not routinely  taken.   Relative
velocity shifts were generally less than 7~\kms.

In 2010 the WIRO-Longslit data were obtained with the 600 l mm$^{-1}$
grating in second order covering 4000--5900~\AA\ at 2.8~\AA\ FWHM
resolution near 5800~\AA.  Reductions followed the same procedures
previously described.  The lower spectral resolution of these data
limited their utility and are seldom used, 
but we note them here for completeness.

We measured the radial velocity for each spectrum obtained at WIRO
using Gaussian fits to the \ion{He}{1} $\lambda$5876 line. Systemic
velocities are, therefore, based solely on this line, by adopting a
rest wavlength of 5875.69~\AA\ measured in model stellar spectra with
static atmospheres \citep[TLUSTY;][]{lanz,hubeny}.  Stars with strong
winds, including very early O stars (not present in this particular
subsample) and evolved stars (MT267, MT734) may exhibit He line
centers that are blue shifted with respect to this assumed rest
wavelength.  Our fitting code\footnote{We use the robust curve
fitting algorithm MPFIT as implemented in IDL \citep{mpfit}.}  fixes
the Gaussian width and depth to be the mean determined from all the
spectra, after rejecting outliers, and it solves for the best fitting
line center.  In the case of an SB2 like MT103, the code fits a
double Gaussian where the widths and depths have been fixed
independently using observations obtained near quadrature.  In the
case of the long-period system MT070 (6.2 years) we included data
from the WIYN telescope \& Hydra fiber spectrograph covering the
range 3800--4500~\AA.  Therefore, we measured radial velocities by
fits to the \ion{He}{1} $\lambda$4471 line instead of $\lambda$5876. 
Our analysis of Keck spectra of stars having minimal radial velocity
variations suggest no systematic differences between these two lines.
However, the velocities from the WIYN/Hydra appear systematically
5--15 \kms\ more negative than velocities from WIRO for
constant-velocity systems having sufficient data from both
observatories to make this comparison.  Observations of radial
velocity standard stars using WIYN/Hydra showed excellent agreement
with published velocities (Paper I).  Given the spectral stability of
the fiber-fed Hydra bench-mounted spectrograph, we consider
velocities from this instrument to be more reliable.  Accordingly,
when WIRO and WIYN spectra are combined to compute solutions for the
same star (MT070), we shift the WIRO velocities by $-$10 \kms. We
caution that this may introduce systematic shifts in the systemic
velocity reported.  However, the primary goal of the Cygnus~OB2
radial velocity survey is to obtain orbital parameters for massive
binaries, a goal that requires good \emph{relative} radial
velocities.  Absolute space velocities from which systemic velocities
may be inferred are of lesser importance.  Tables~\ref{103.tab} and
\ref{rest.tab} report the adopted radial velocities and uncertainties
for each star, along with the orbital phase, $\phi$, from the
best-fitting solution, and the observed minus computed (O$-$C)
residuals.  The final column specifies the observatory used to obtain
the data.

\section{Orbital Solutions}
 
We analyzed the CLEANed \citep{Roberts87} power spectrum for each
object to select likely periods and then examined the folded velocity
curve for periods corresponding to the strongest peaks in the power
spectrum.  In most cases the strongest peaks yielded clean,
convincing phased velocity curves.  Secondary peaks and possible
aliases could be eliminated by visual inspection owing to the much
larger dispersion in the data at any given phase.  We used the binary
orbital solution package ``binary'' by D.
Gudehus\footnote{http://www.chara.gsu.edu/~gudehus/binary.html} with
these initial period estimates and the radial velocity data to solve
for the full suite of orbital parameters and associated
uncertainties.  Table~\ref{solutions.tab} compiles these best-fitting
parameters and uncertainties for each object.  Listed within the
table are the period in days ($P$), eccentricity of the orbit ($e$),
longitude of periastron in degrees ($\omega$), systemic radial
velocity ($\gamma$), epoch of periastron ($T_0$), primary and
secondary (for MT103) projected semi-amplitudes  ($K_1$ \&\ $K_2$),
spectral classifications from this survey (S.C.$_1$ \&\ S.C.$_2$),
estimates of the mass ratio ($q$), inclination ($i$), semi-major axis
($a$), and reduced chi squared values of the best fitting solution. 
The ensuing subsections elaborate on the detailed orbital solutions
for the five massive systems.

\subsection{MT070}

MT070 has $V=12.99$ with a systemic velocity of $-$6.4 km/s, as
reported by \citet{Kiminki07}.  Its $B-V$ color of 2.10~mag makes it
among the reddest early-type stars in Cyg~OB2.  Figure~\ref{color}
show that MT070 lies along the eastern edge of the Association (upper
right) in an apparently unremarkable region mostly devoid of H$\alpha$
and PAH emission.  However, there are filaments of 8 $\mu$m \& 24
$\mu$m emission and \co\ (C. Brunt, private communication) at Cyg~OB2
radial velocities cutting across this region, suggesting the
likelihood of elevated dust extinction that could explain the redder
color of MT070 relative to the other Cyg~OB2 members.

\begin{figure}
\epsscale{1.0}
\centering
\plotone{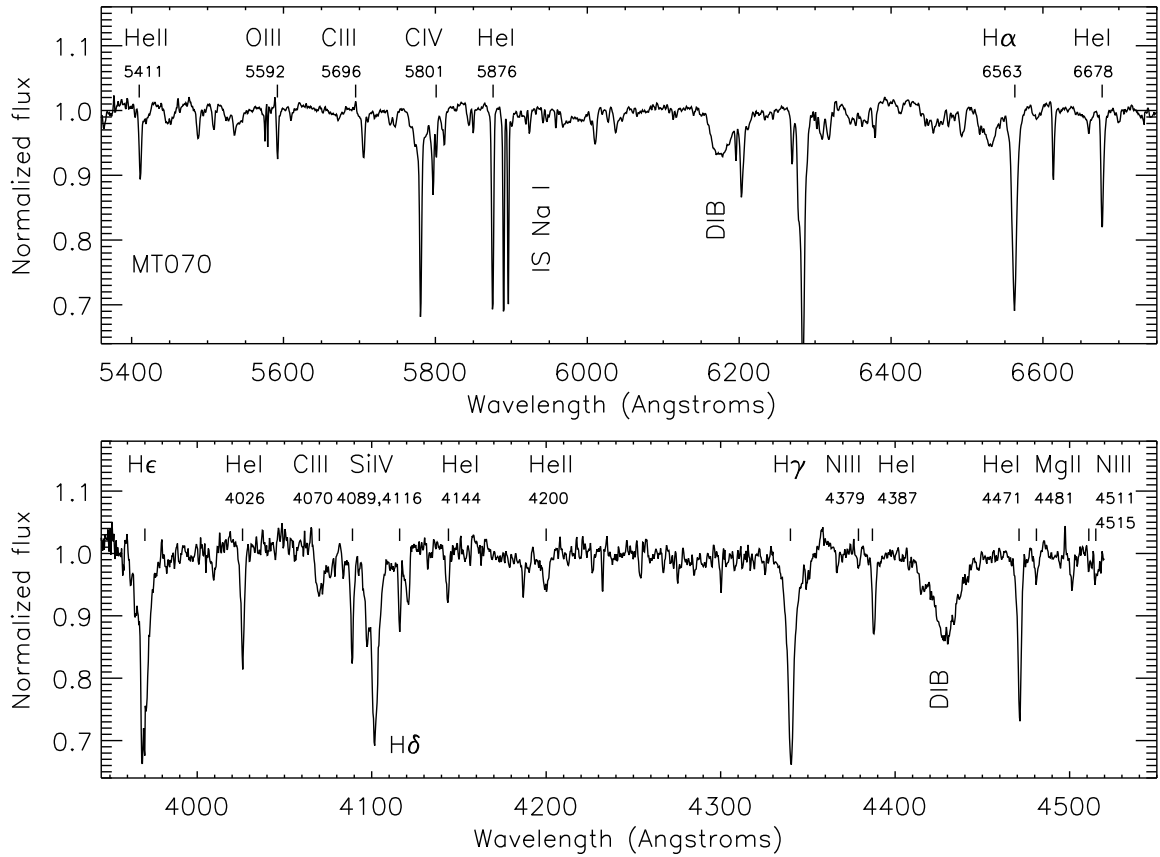}
\caption{Average spectrum of MT070 using data from WIRO (top) and WIYN (bottom).  
Tickmarks and labels mark key spectral features. Most of the unlabeled absorption features in
the upper panel are from diffuse interstellar bands \citep{jenniskens}.
\label{070spec}}
\end{figure}

Figure~\ref{070spec} displays an average of ten WIRO spectra of MT070
(top) covering the red spectral range and six WIYN spectra (bottom)
covering the blue spectral range.  Labels denote key spectral
features.  The composite WIYN spectrum for MT070 shows \ion{He}{2}
$\lambda$4200:\ion{He}{1} $\lambda$4144, one of the primary
temperature-sensitive line ratios in this wavelength regime, in a
ratio of 1:1, indicating the temperature class of the primary is
approximately O9\citep{Sota2011}, in agreement with the initial
estimate of \citet{Kiminki07}.  Additionally, the luminosity-sensitive
ratio \ion{Si}{4} $\lambda$4089:\ion{He}{1} $\lambda$4026 appears in a
ratio of approximately 3:4, indicating an evolved luminosity class of
III--II \citep{Sota2011}. In the composite WIRO spectrum, \ion{C}{3}
$\lambda$5696 appears to be slightly in emission,
\ion{O}{3}~$\lambda$5592:\ion{He}{2} $\lambda$5411 appears in a ratio
of approximately 3:5, and \ion{C}{4} $\lambda$5801:\ion{He}{2}
$\lambda$5411 appears in a ratio of approximately 1:2, also indicating
an evolved type \citep{Walborn2009}. We therefore adopt the spectral
type of O9III for the primary of MT070.  This is consistent with the
relative faintness ($V=$12.99) but the larger extinction of $\Delta
A_V$=$+$1.8 mag (inferred from $B-V$=2.10 mag) relative the average
Cyg~OB2 member ($B-V$$\simeq$ 1.5 mag).

\begin{figure}
\epsscale{1.0}
\centering
\plotone{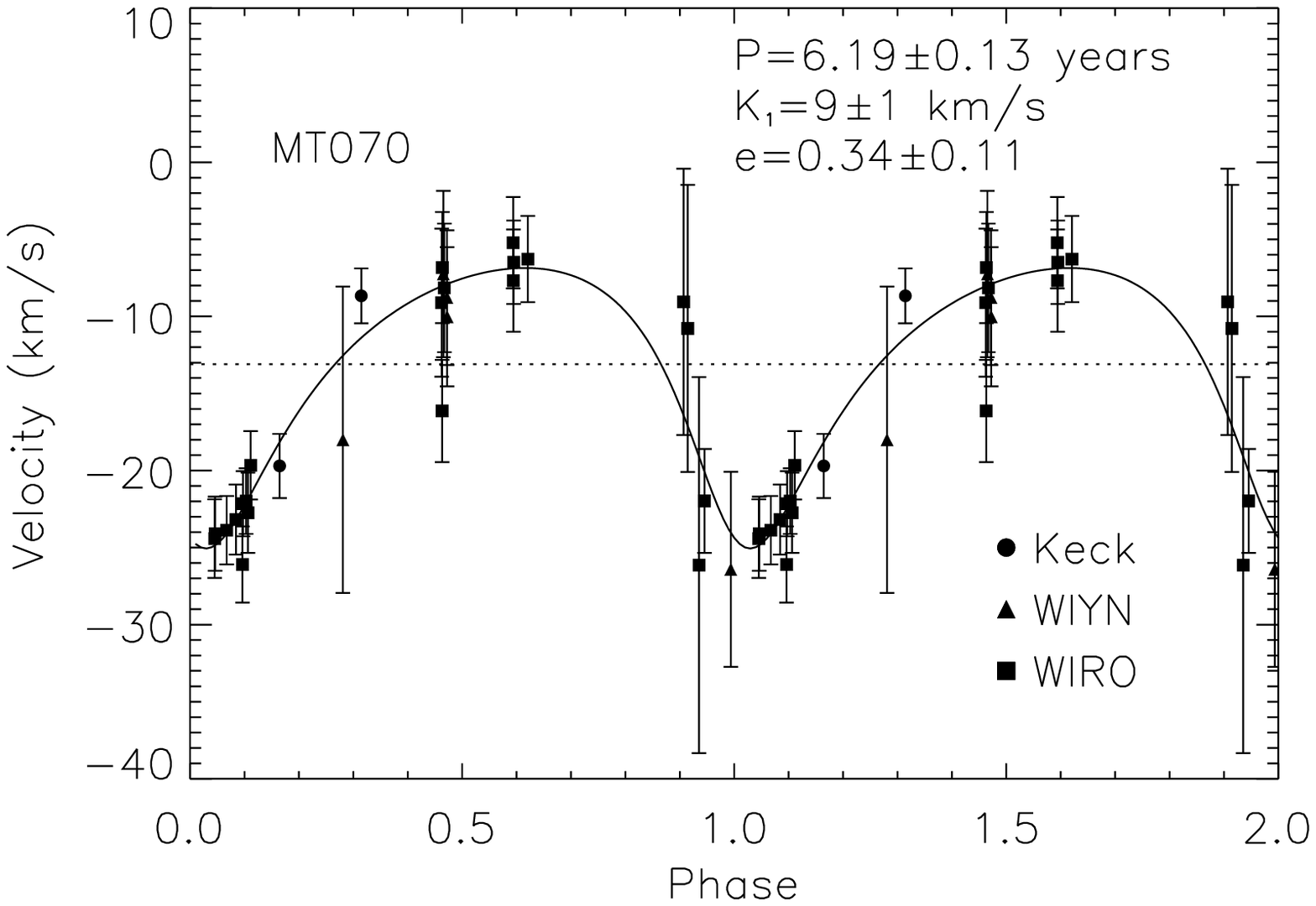}
\caption{Radial velocity data and best-fit orbital solution curve 
for the 6.19 year
SB1 system MT070 (O9III:) using 21 measurements from WIRO (filled squares), 
five from WIYN (triangles), and two from Keck (circles).
The four points with the largest error bars are from the 2010 WIRO campaign
which suffered from low spectral resolution.
\label{070curve}}
\end{figure}

Our spectroscopic dataset on MT070 includes 28 epochs covering the
period 1999--2011 from the Keck (HIRES spectrograph; 2 measurements),
WIYN (Hydra spectrograph; 5 measurements), and WIRO (WIRO-Longslit
spectrograph; 21 measurements) observatories.  Our time-series
analysis of the MT070 spectra shows a peak in the power spectrum near
6 years, along with several secondary peaks at similarly long periods
between 5--8 years.  However, the best-fitting orbital solutions
invariably converged to a period of 6.19$\pm$0.13 years, which yielded
$\chi_{red}^2$=0.68, considerably smaller than other candidate
periods. Figure~\ref{070curve} shows the data and best fitting orbital
solution.  Filled circles, triangles, and squares denote the Keck,
WIYN, and WIRO data, respectively.  This solution specifies a
systemtic velocity $\gamma$=$-$13$\pm$1 \kms, eccentricity
$e$=0.34$\pm$0.11, and a (projected) primary velocity amplitude
$K_1$=9$\pm$1 \kms.  Table~\ref{solutions.tab} lists the full orbital
parameters for MT070 and other systems. The four data points near
phase $\phi=0.9$ are from WIRO in 2010, and, despite the larger
uncertainties, these are included to constrain a portion of the orbit
otherwise lacking data. The radial velocities reported in
Table~\ref{rest.tab} for data from WIRO have been corrected by $-$10
\kms\ to bring it into better agreement with the WIYN data.

If we adopt a primary mass of $M_1$=22 \msun\ based on the O9III
spectral type \citep{martins05}, we can begin to place constraints on
the secondary mass $M_2$ and on the inclination of the system.  If the
inclination of the system were as low as $i$=18\degr, the implied
secondary mass approaches that of the primary.  Making the motivated
assumption that the secondary is on the main sequence, the luminosity
of such a star would be smaller than that of the primary, and indeed,
we do not see spectral features from the secondary.  Therefore, we can
limit the inclination to $i>18$\degr, making the mass ratio
$q=M_2/M_1<1$ and semi-major axis $a<10.9$ AU.  Stronger constraints
are not possible because of the low velocity amplitude of the system
relative to the spectral resolution of the data.  On the other
extreme, if $i=90^\circ$, the implied secondary mass is 5.0 \msun\
(approximately a B6V), $q$=0.22, and $a$=9.3 AU.  Photometric surveys
of Cyg~OB2 \citep{PJ98,Henderson2011} do not list MT070 as an
eclipsing system, but, given the long period, the lack of observed
eclipses does not place useful constraints on the inclination of the
system.

For completeness, we also report briefly the orbital solution  that
would have resulted without applying the -10 \kms\  systematic
velocity correction to the WIRO data. In this case, we find a a
similar period but a much more eccentric system:  $P=5.79\pm0.11$~yr,
$e=0.71\pm0.22$, a systemic velocity $\gamma=-1\pm4$ \kms,   and a
velocity amplitude $K_1$=24.4 \kms. The $\chi^2_{red}$ is also
considerably larger at 1.12.  Consequently, we adopt the 6.19~yr
period and solution described previously. In summary, MT070 is a
single-lined long-period binary with $P=6.19$~yr, $e\simeq0.34$,
$0.22<q<1$, and $18^\circ<i<90^\circ$. 

\subsection{MT103}

MT103 lies just 195\arcsec\ from MT070, and it is similarly red, with
$B-V$=2.00. It is the only double-lined binary reported in this work,
but many SB2 systems within Cyg~OB2 were described in Papers
II--IV.   The components are significantly blended, except for the
times very near quadrature.  We determine velocities using double
Guassian fits to the \ion{He}{1} $\lambda5876$ line for the 12 WIRO
data (all from 2011) and  $\lambda$4471 for the three WIYN data (two
from 2001 and one from 2006).  We chose to  constrain the  width of
the lines (we used a FWHM of 3.0~\AA, estimated from the most
deblended spectra) and the ratio of  component line depths to be near
2.5:1, within 20\%.   The power spectrum of the primary exhibited a
singular strong peak near 22~days with no credible aliases.  The
power spectrum of the secondary  was consistent with this value but
considerably more noisy owing to larger velocity uncertainties.   As
with MT070, we applied a $-$10 \kms\ shift to all of the WIRO data to
improve agreement with the WIYN velocity scale.         

\begin{figure}
\epsscale{1.0}
\centering
\plotone{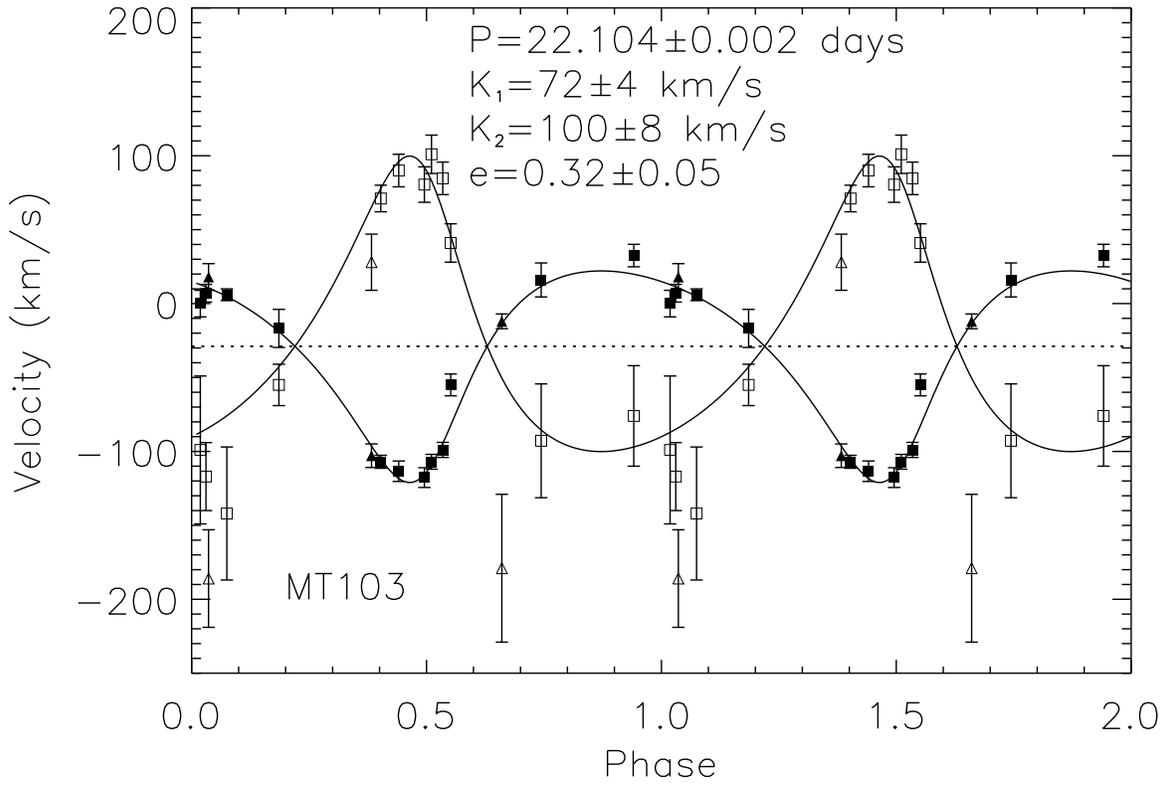}
\caption{Radial velocity data and best-fit orbital solution curve 
for the 22.10 day SB2 system MT103.  
Filled symbols denote the primary and open symbols show the secondary.
Triangles show WIYN data and squares show WIRO data.  
\label{103curve}}
\end{figure}

Figure~\ref{103curve} shows the best-fitting orbital solution
performed jointly for the primary and secondary stars.   The period
of 22.104$\pm$0.002~days is quite secure, anchored by the large time
baseline of the combined WIYN and WIRO data.  The eccentricity of 
$e=0.32\pm0.05$ is likely to be less secure, owing to the difficulty
in resolving the blended spectral features.  The velocity
semi-amplitudes of  $K_1$=71$\pm$4 \kms\ and $K_2$=100$\pm$8 \kms\
for the primary and secondary, respectively,  indicate a mass ratio
of 0.71$\pm$0.10.  The velocity data for the secondary has large
uncertainties, particularly near $\phi=$0.6--1 where the deblending
technique is only marginally effective.  Table~\ref{solutions.tab}
includes the full list of orbital elements.

\begin{figure}
\epsscale{1.0}
\centering
\plotone{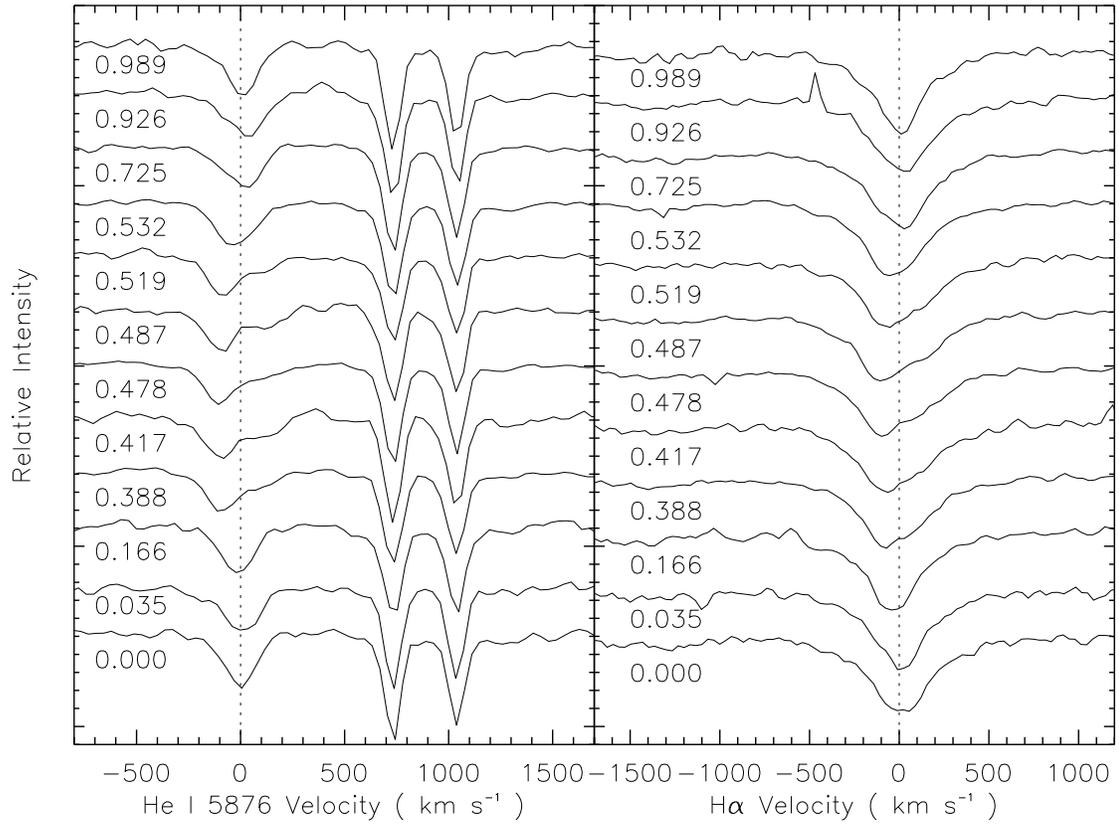}
\caption{A sequence of the 12 WIRO spectra for MT103, ordered by phase, in
the spectral region around \ion{He}{1} $\lambda$5876 (left) and H$\alpha$ (right).  The Na~D doublet lies
to the right of the He $\lambda$5876 line.   \label{103phase}}
\end{figure}

Figure~\ref{103phase} (left panel) displays a portion of the 12
spectra around \ion{He}{1} $\lambda$5876 from the 2011 WIRO observing
campaign, ordered by phase.  The right panel shows the portion of the
spectra surrounding H$\alpha$.  The separation of the components
begins near $\phi$=0.16 as the primary becomes blueshifted, and the
line separation is well-developed  at the level of $\sim$200 \kms\ at
$0.38<\phi<0.53$.   Near $\phi=0.72$ the primary becomes redshifted
and the secondary's line is visible on the blueshifted wing until
$\phi\simeq0.95$ or later.  

\citet{Kiminki07} list MT103 as a B1V. A qualitative analysis of the
composite of 3 WIYN spectra covering the blue wavelength range, using
\citet{Gray2009}, supports this classification. In the composite
spectrum, the \ion{Si}{4} $\lambda\lambda$4089,4116 lines are
absent. This suggests that the primary component is a B1 or
later. \ion{Mg}{2} $\lambda$4481 is still very weak
relative to \ion{He}{1} $\lambda$4471, and \ion{Si}{2}
$\lambda$4128--30 is also very weak, suggesting that it is earlier
than B3. The absence of \ion{O}{2} $\lambda$4070, \ion{O}{2}
$\lambda$4348, \ion{O}{2} $\lambda$4416 and \ion{N}{2} $\lambda$3995,
which we acknowledge may be lost in the noise of the spectral
continua, indicates that the primary is not evolved, and lies in the
range B1--2V.  We adopt B1V for the primary. The most deblended
spectra indicate that the secondary component's He lines are similar
to the primary's, but slightly weaker. Given this and that the
calculated mass ratio for this system is 0.71, we adopt B2 as the
likely spectral type for the secondary.  Our attempt at a spectral
synthesis of a B1V (14.2 \msun; $M_V$=$-$3.2)\footnote{Interpolated
from \citet{drilling}.} and a B2V (10.9 \msun; $M_V$=$-$2.45) provides
reasonable agreement with the most deblended spectra and is consistent
with the measured mass ratio. This suggests a luminosity ratio of
about 2:1 and a mass ratio consistent with the measured 0.71 value.
These masses imply an inclination of $i\simeq50$\degr\ in order to
match the observed radial velocities.  The resulting semi-major axis
of the systems is then $a\simeq$0.45~AU.

The measured systemic velocity of $-$28$\pm$3 \kms\ is noteworthy for
being about 15 \kms\ more negative than the mean of $V_{helio}$=$-$13
\kms\ for massive stars in Cyg~OB2 \citep{Kiminki07}. Without the
10 \kms\ velocity offset applied to the WIRO data, the systemic
velocity for MT103 would be in better agreement with the rest of
Cyg~OB2 massive stars.  We note here that the uncertainty on the
systemic velocity is likely to be somewhat larger than the formal 3
\kms\ error quoted and that we do not consider the difference from
$-$13 \kms\ to be  significant, given possible systematic velocity
uncertainties at the 10--12 \kms\ level.

\subsection{MT174}

\begin{figure}
\epsscale{1.0}
\centering
\plotone{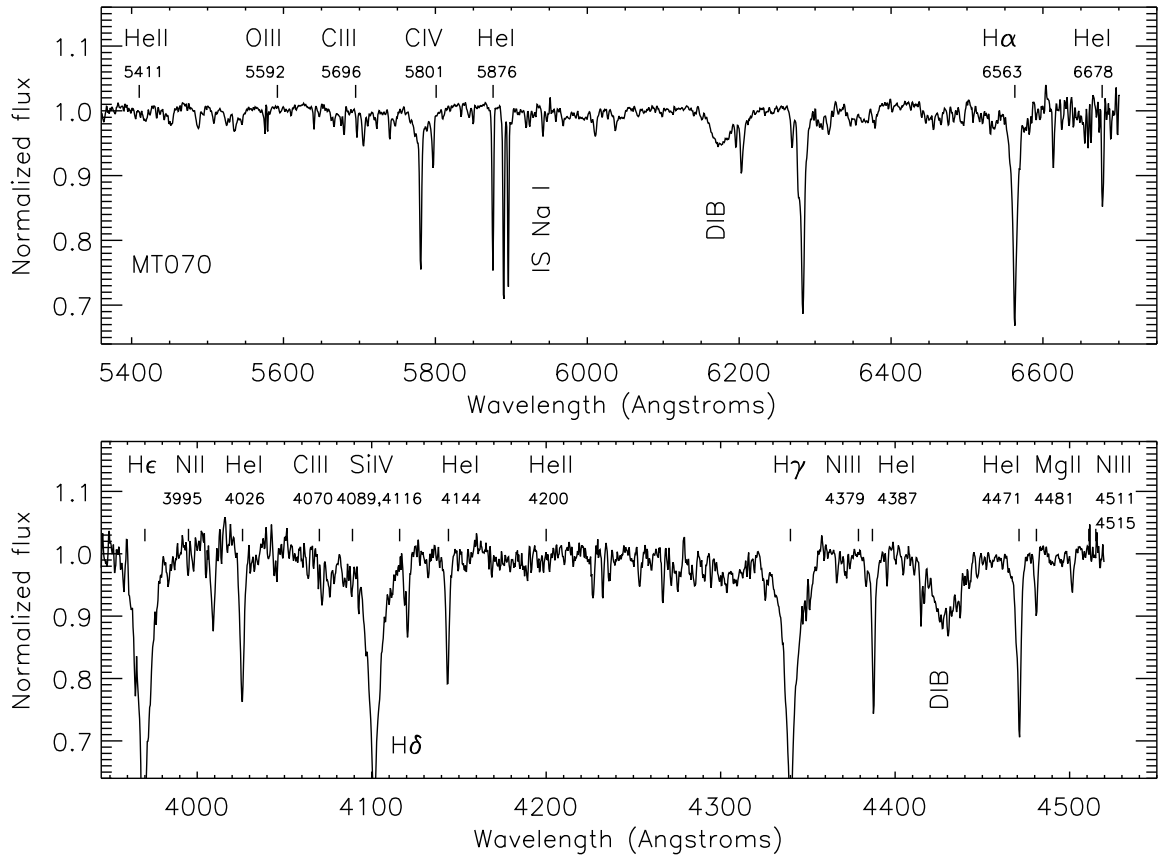}
\caption{Average spectrum of MT174 (B2III) using data from WIRO (top) and WIYN (bottom).  
 \label{174spec} }
\end{figure}

\citet{Kiminki07} classify MT174 as a B2IV, while \citet{MT91}
estimated B2V.  Figure~\ref{174spec} displays an average spectrum of
MT174 using ten 2011 WIRO data (top) covering the red spectral range
and using seven WIYN spectra (bottom) covering the blue spectral
range.  Labels mark key spectral features.  The primary
temperature-sensitive line ratio in this wavelength regime is
\ion{Si}{4} $\lambda$4089:\ion{Si}{3} $\lambda$4552 \citet{Gray2009}.
Unfortunately, the composite WIYN spectrum does not cover \ion{Si}{3}
$\lambda$4552.  However, the spectrum does show very weak \ion{Si}{4}
$\lambda$4089 and no \ion{Si}{4} $\lambda$4116, suggesting a
temperature class of B1--B2. The \ion{He}{1} $\lambda$4471:\ion{Mg}{2}
$\lambda$4481 temperature-sensitive ratio, which becomes useful around
B2--B3 is $\sim$1:5 and indicates a temperature class of B2 or
earlier \citep{Walborn90}. Additionally, \ion{He}{1} $\lambda$4009 and
\ion{He}{1} $\lambda$4026 are quite strong and in a ratio of 1:2
respectively, also indicating a temperature class of approximately
B2 \citep{Walborn90}. The composite WIYN and WIRO spectra show narrow Balmer
line widths. In conjunction with the weak, but clearly present
\ion{N}{2} $\lambda$3995 and \ion{O}{2} $\lambda\lambda$4070,4076,
this suggests a luminosity class of III.  Finally, a direct comparison
to HD35468 (B2III) from the \citet{Walborn90} digital atlas provides
the best agreement with both the composite WIYN and WIRO spectra.
Therefore, we adopt a spectral type of B2III for the primary component
of MT174. At V=12.55 and B-V=1.21, it is not unusually colored for
early type stars in Cyg~OB2, but it is 1--2 magnitudes brighter than
other B2 stars in our CygOB2 sample, consistent with an evolved
luminosity class.

\begin{figure}
\epsscale{1.0}
\centering
\plotone{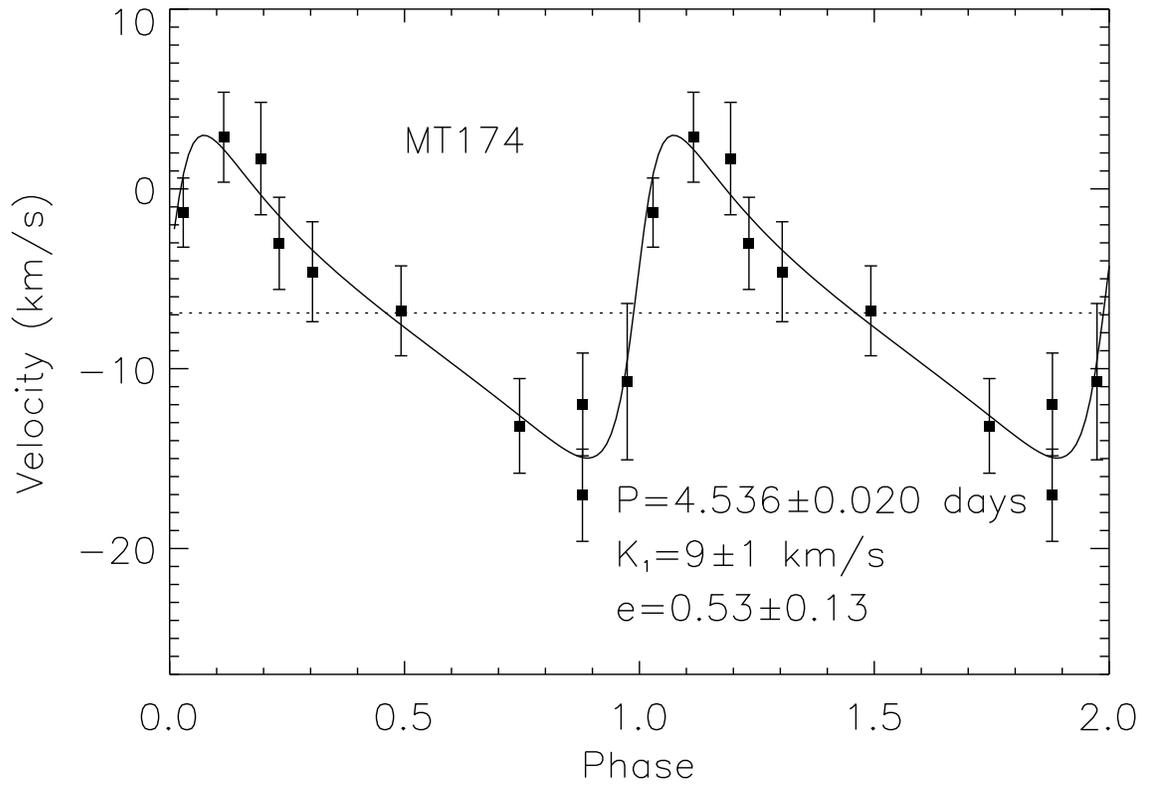}
\caption{Orbital solution for the SB1 system MT174 (B2III) using 10 WIRO data from
2011.  
 \label{174curve} }
\end{figure}

Our 10 WIRO spectra from 2011 reveal that this is a single-lined
spectroscopic binary with a period of 4.536$\pm$0.020 d,
$e=0.53\pm0.13$, and  $K_1$=9$\pm$1 \kms.  
Figure~\ref{174curve} shows the data and best fitting orbital
solution, which has $\chi^2_{red}$=0.71.  The systemic velocity of
+11$\pm$2 \kms\ is considerably more positive than the $-$13 \kms\
average of probable Cyg~OB2 members \citep{Kiminki07}.  This may be
the result of systematic  differences between the WIRO data and that
from other observatories.   Our attempts to include the seven
measurements from WIYN over the period 2001 -- 2006 resulted in poor
or non-convergent solutions, even after applying a $-$10 -- $-$20 \kms\
offset to the WIRO measurements.  This may indicate the presence of
additional kinematic variations in the MT174 system, possibly the
result of the gravitational influence of a third body on timescales
of many months or years.  Unfortunately, our most reliable data do not have
sufficient time coverage to constrain the nature of possible 
additional velocity perturbations in MT174. Given that the mean
velocity measured in our WIYN spectra  is $V_{helio}$=$-$6 \kms\
compared to +12 for the WIRO data, we apply a $-$18 \kms\ zero-point
shift to our final orbital solution, summarized in
Table~\ref{solutions.tab} and the radial velocities in
Table~\ref{rest.tab}.  The resulting systemic velocity of -6 \kms\ is
more consistent with the ensemble of Cyg~OB2 stars.   

If we adopt a primary mass appropriate to a B2III  ($M\simeq$10.9
\msun; $R$=10 $R_\odot$), the implied secondary mass is $M_2$=0.35
\msun\ for $i=$90\degr\ and $M_2$=1.0 \msun\ for $i=$20\degr.  For
inclinations as low as 3\degr\ $M_2$ approaches $M_1$, a lower bound
imposed by the lack of secondary spectral features.  The absence of
known eclipses \citep{Henderson2011}, coupled with  the narrow
linewidth of $\sim$1~\AA\ FWHM for \ion{He}{1} $\lambda$4471 in our
single 1999 Keck spectrum, is consistent with a small projected
rotational velocity, which would imply a small inclination angle if
the components' spin axes are aligned with the orbital axis.  (The
small linewidth is also consistent with  the evolved luminosity
classification.)  The inclination is, effectively, unconstrained, and
may lie in the range $3^\circ < i < 87^\circ$.  For all inclinations 
greater than about 15 degrees, the secondary is M$<$1.4 \msun, making
MT174 one of the most extreme mass ratio ($q<0.13$)  systems yet
uncovered in Cyg~OB2.  

MT174 is an interesting system in that it is the first binary in the
Cyg~OB2 radial velocity survey to contain a massive star and a
quasi solar-mass star.  As such, it is a potential progenitor of a
low-mass X-ray binary system (LMXRB).   The distribution of mass
ratios among Cyg~OB2 stars  is approximately flat, and some 45\% of
massive binary systems have components capable of mass transfer,
\citep{Kiminki2012b}, meaning that $\sim$5--10\% of massive binaries may
be progenitors of low-mass X-ray binaries. However, the  relative
paucity of observed LMXRBs in nature implies that the formation
channels for such objects are narrow and require special conditions
during post-main-sequence evolution, as discussed by
\citet{vanden83}.

\subsection{MT267 (A11)}

\begin{figure}
\epsscale{1.0}
\centering
\plotone{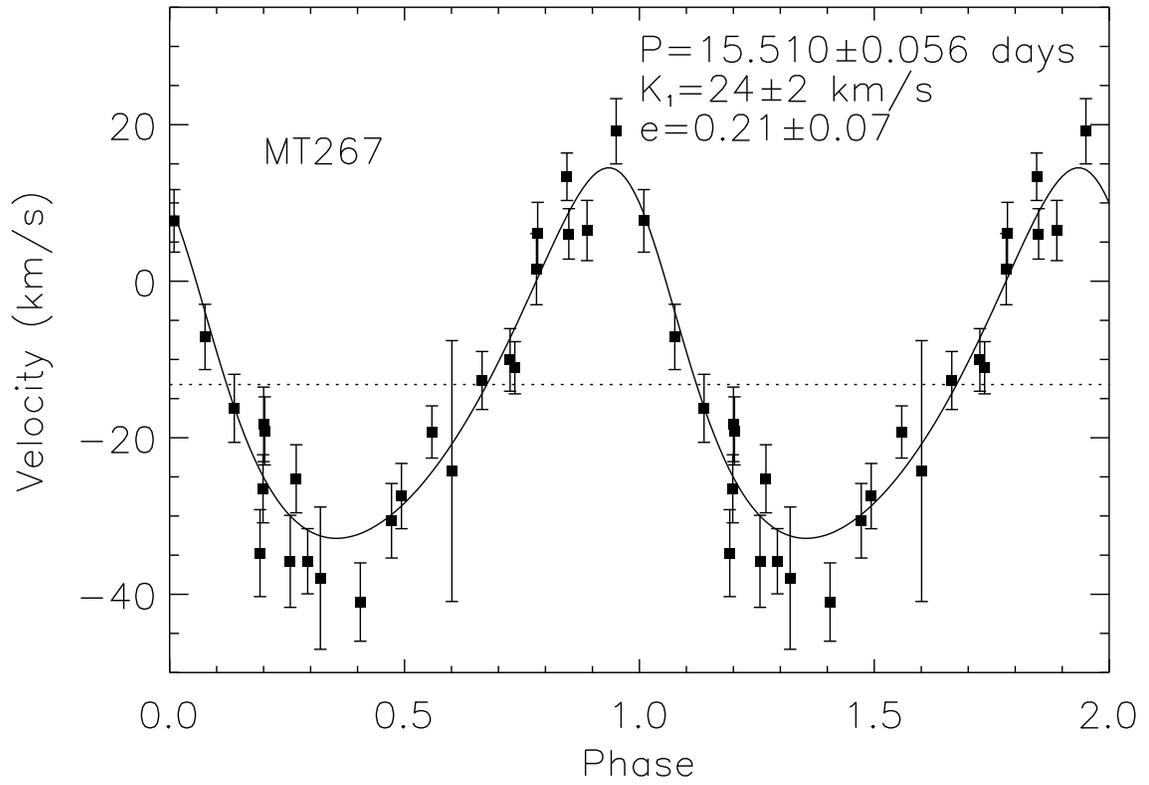}
\caption{Orbital solution for the SB1 system MT267 (O7.5III--I). 
\label{267curve}}
\end{figure}

Figure~\ref{267curve} shows the 25 WIRO data, all from 2011 June --
2011 November, and the orbital solution for the SB1 system MT267 (A11
in the notation of \citealt[][]{comeron02}).   The period of
$P=$15.511$\pm$0.056 d, eccentricity of $e=$0.21$\pm$0.07, and velocity
amplitude $K_1$=24$\pm$2 \kms\ are well constrained by the data. 
However, the $\chi^2_{red}$ of 1.59 is larger than for most
other systems, suggesting the possibility of photospheric line
variations (common for evolved massive stars) or an additional
dynamical influence in the system.  Our examination of the  observed
minus computed (O-C) values in Table~\ref{rest.tab} reveals a
weak correlation between Julian date and O-C, but additional
data will be needed to establish the nature of additional long-term
velocity variations, such as those expected from a third body.   The 
resulting systemic velocity of $\gamma$=$-$13$\pm$1 \kms\  is consistent
with other CygOB2 stars, and, given the lack of data from other
observatories, we do not apply any systematic correction to the
velocities plotted in Figure~\ref{267curve} or the data in
Table~\ref{rest.tab}.  The full suite of orbital parameters appears
in Table~\ref{solutions.tab}.

\begin{figure}
\epsscale{1.0}
\centering
\plotone{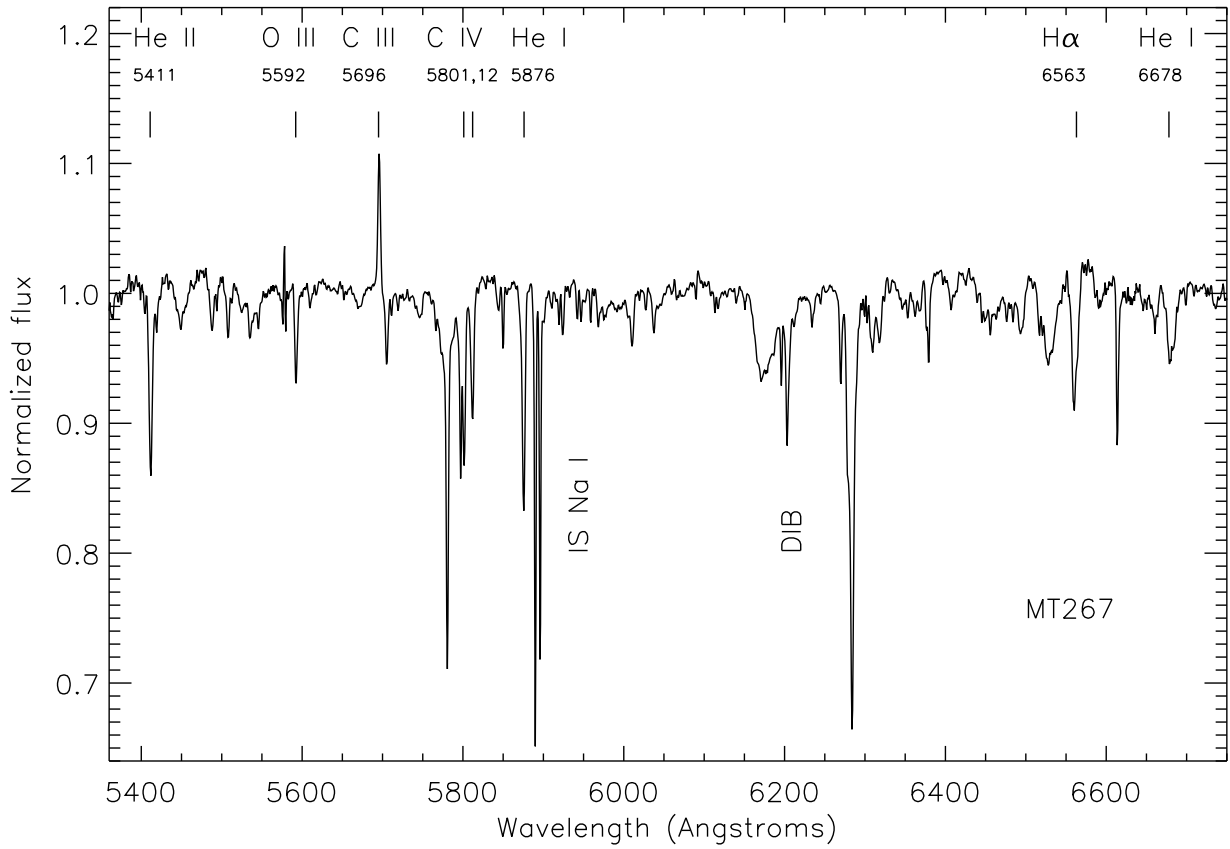}
\caption{Average of 25 WIRO spectra for MT267. Labels mark key
spectroscopic features of hot stars along with some prominent insterstellar
features.  Most of the unlabeled absorption features 
 are from diffuse interstellar bands \citep{jenniskens}.
\label{267spec}}
\end{figure}

\begin{figure}
\epsscale{1.0}
\centering
\plotone{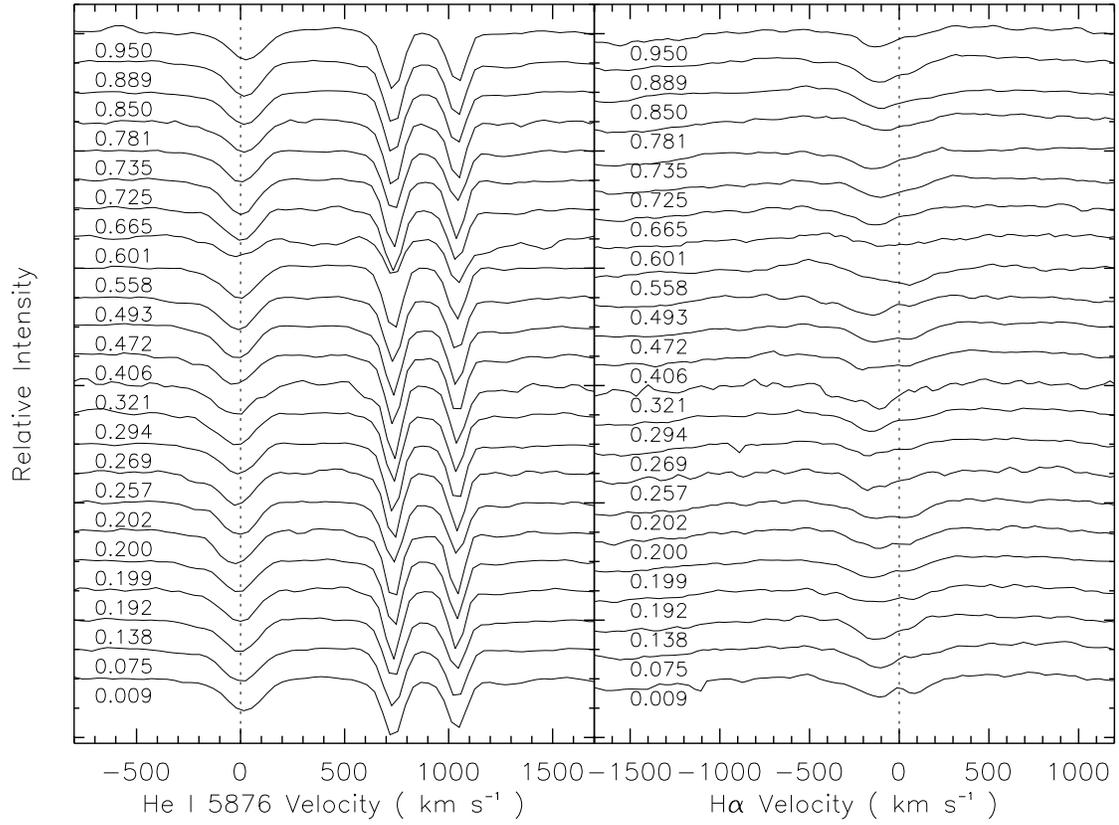}
\caption{A sequence of spectra covering H$\alpha$ and \ion{He}{1} $\lambda$5876
for MT267, ordered by phase.  This O7III: star exhibits variability in the H$\alpha$
line profile, but not in the He line profiles.
\label{267sequence}}
\end{figure}

Figure~\ref{267spec} is an average spectrum of MT267 covering the
wavelength range 5400--6700~\AA\ from our WIRO data. Labels mark key
spectral atmospheric and interstellar medium features. The composite
spectrum exhibits a \ion{He}{2} $\lambda$5411:\ion{He}{1}
$\lambda$5876 ratio of near unity, indicating a temperature class near
O7 \citep{Walborn80}. \ion{O}{3} $\lambda$5592 is present at roughly
1:3 with \ion{He}{1} $\lambda$5876 and also agrees with a late-O
temperature class \citep{Walborn80}.  \ion{C}{3} $\lambda$5696 is
prominent in emission.  \ion{C}{4} $\lambda\lambda$5801--5812 and
\ion{O}{3} $\lambda$5592 are present at moderate strength and
relatively broad, and H$\alpha$ appears to be partially filled in by
emission. This all indicates an evolved star \citep{Walborn2009,
Walborn80}.  H$\alpha$ shows variable emission, which is also
indicative of an evolved star or an interacting
binary. Figure~\ref{267sequence} displays a sequence of H$\alpha$
(right) and \ion{He}{1} $\lambda$5876 (left) spectra, ordered by
phase. Changes in the H$\alpha$ line profile are evident, but
\ion{He}{1} appears unchanged, except for the periodic velocity
variations. Comparing spectral standards from the \citet{jacoby84}
digital atlas with a composite spectrum of this star, we find that an
O7--O8 best fits the temperature of primary component, with the
luminosity class in the range III--I.  This also agrees with our
photometric assessment of this star and is consistent with the
spectral type estimated by \citet{Negueruela08} who give
O7.5Ib--II(f).

If we adopt a mass of 32 \msun\ appropriate for an O7.5III -- I
\citep{martins05} then the implied secondary mass ranges between 3.0
\msun\ for an inclination of $i=90$\degr\ and 30 \msun\ for
$i=8$\degr\ if the secondary approaches the mass of the primary.  
Given the evolved, luminous nature of the primary, which we assume
dominates the luminosity of the system, is it conceivable that the
secondary may be as massive as the primary but not visible in the
integrated spectrum.  The lack of known eclipses
\citep{Henderson2011} provides only a weak constraint that the
inclination is smaller than 81\degr.   Thus, we conclude that the
mass ratio of the system is $0.09<q<1$ with semi major axis
$0.48>a>0.39$ AU.  The presence of X-rays and X-ray variability in
MT267 \citep{rauw2011} is consistent with a wind-wind collision in a
close, interacting binary.   If the wind collision scenario is
correct for the origin of the X-rays, then the secondary component is
probably toward the massive end of our allowed range,  implying
$i<13$\degr\ in order to make the mass of the secondary equivalent to
a B0V or earlier.

\begin{figure}
\epsscale{1.0}
\centering
\plotone{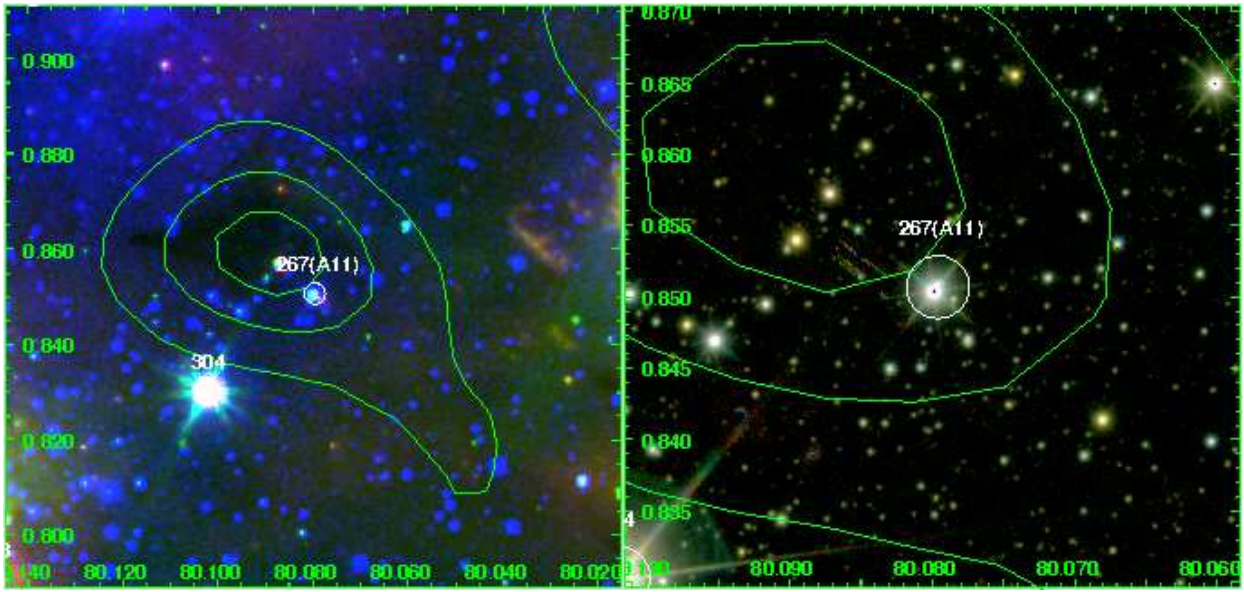}
\caption{The field around MT267.  The left panel shows the optical/mid-IR view
with POSS R-band in blue and {\it Spitzer} GLIMPSE 8 $\mu$m and 24 $\mu$m in green and red, respectively.
Green contours depict the \co\ globule, which lies coincident with an IR dark cloud.
The right panel shows the UKIDSS JHK appearance (blue/green/red) of the field at a higher zoom factor.
\label{267environ}}
\end{figure}

MT267 has an unusually red $B-V$ color (2.19 mag) compared to other
massive stars in Cyg~OB2 ($<$B-V$>$$\simeq$1.4; \citealt[][]{MT91}),
meaning a higher foreground dust column density toward this star.
Figure~\ref{267environ} shows an infrared view of the field around
MT267.  The left panel is a three-color image with blue/green/red
representing the POSS R-band, the {\it Spitzer} 4.5 $\mu$m band, and
{\it Spitzer} 24 $\mu$m band.  In the right panel blue/green/red
represent the UKIDSS $JHK$ images.  Labels mark the locations of MT267,
and the well-known, heavily reddened mid-B supergiant Cyg~OB2 \#12
(MT304).  Interestingly, these two evolved massive stars lie within a
radius of 53\arcsec\ from each other.  Clearly evident in the left
panel is the infrared dark cloud adjacent to both objects. This
is the Bok globule noted by \citet{scappini} and \citet{poppi}.  Green
contours show the \co\ line intensity integrated over the velocity
range $V_{LSR}$=9.7 -- 11.7 \kms, corresponding to $V_{\odot}$=23.7 --
25.7 \kms.\footnote{The conversion from the local standard of rest
(LSR) to heliocentric reference frame is $V_\odot = V_{LSR}$ +14 \kms\
in the direction of CygOB2.}  Contour levels show 13, 25, 38, and 50 K
\kms\ intensities corresponding to column densities of $N(H_2)$=3.9, 7.5, 11.4,
and 15.0$\times10^{21}$ cm$^{-2}$, using the mean relation between
molecular column density and \co\ integrated intensity
$a=3.0\times10^{20}$ cm$^{2}$/(K~km~s$^{-1}$) \citep{solomon87}.
Given the conversion factor of \citet{bohlin78}\footnote{ ${A_V }\over{R_V}$ =
$<E(B-V)>$ (mag) = $N(H+H_2)$ cm$^{2}$ / $5.8\times10^{21}$ (cm$^{2}$
mag$^{-1}$ )}, the contours correspond roughly to extinctions
$A_V$=0.67, 1.3, 2.0, 2.6 mag, assuming $R_V=3.1$.  These are probably
lower limits given that \co\ becomes optically thick and/or freezes
onto dust grains in the cold cores of molecular clouds.  If this Bok
globule/molecular cloud lies in the foreground to MT267 (and Cyg~OB2\#12),
it is conceivably responsible for the higher extinction recorded to
these objects.  However, the very large extinction of Cyg~OB2\#12
($A_V\sim$12 mag) cannot be solely attributed to this cloud.  Given
the large radial velocity difference between this cloud and Cyg~OB2,
it is probable that this cloud lies substantially in the
foreground. The lack of PAH emission around the borders of this cloud
further supports the conjecture that it lies at some distance from the
massive ionizing stars.

\subsection{MT734 (Schulte \#11)}

\begin{figure}
\epsscale{1.0}
\centering
\plotone{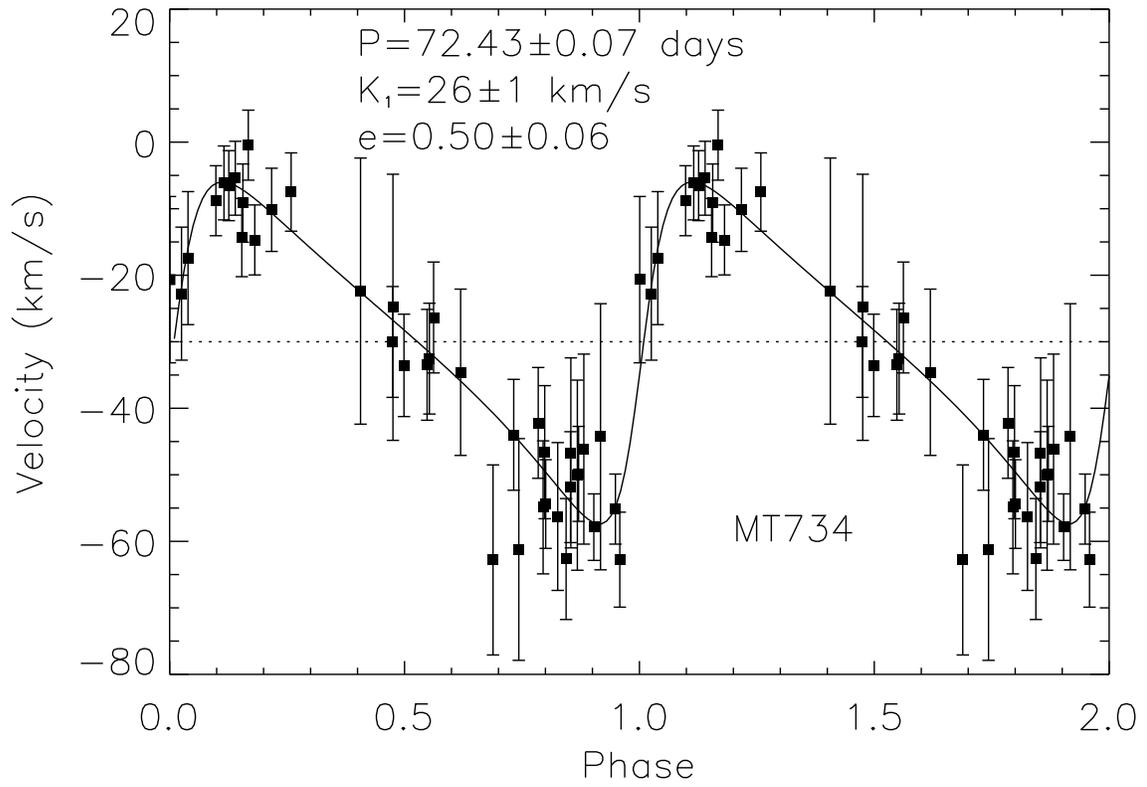}
\caption{Orbital solution for the O5I(f) SB1 system MT734 based on 39 WIRO spectra from 2007--2011. 
\label{734curve}}
\end{figure}

MT734 (VI~Cyg~\#11/Schulte~11) is a well-known emission-line star of
type O5I(f), as classified by \citet{herrero2002}.  
Figure~\ref{color} illustrates that MT734 lies along the western
boundary of the Association in an area apparently devoid of H$\alpha$
and PAH emission, although it is surrounded by molecular clouds. Our
39 spectra, taken with the WIRO Longlit spectrograph over the period 
2007--2011, show that this system is a single-lined spectroscopic
binary.  We find a period of 72.43$\pm$0.07 d, $e=$0.50$\pm$0.06, and
$K_1$=26$\pm$1 \kms.  Figure~\ref{734curve} depicts our best
orbital solution for this system.  MT734 possesses a systemic
velocity of  $-$31$\pm$1 \kms\, which is more negative than most
members of the Association.  We have not applied any systematic
velocity corrections to the WIRO data.  The comparatively
blue-shifted  line centers may be a result of line formation in the
upper atmosphere/outflowing wind in this very luminous
supergiant.    

If we adopt a mass of 51 \msun\ appropriate to an O5I
($R=19~R_\odot$) from \citet{martins05}, then an inclination of 90\degr\
yields a secondary mass of $M_2=$7.7 \msun, $q$=0.15, and $a$=1.31 AU.
Assuming $i$=13\degr\ implies that the seondary mass approaches
that of the primary, $q\simeq$~1 and $a$=1.51 AU.  The lack of known
eclipses does not place helpful constraints on the
inclination.  MT734 is listed as a 1400 MHz radio continuum source
with $S_{1400}$=0.5$\pm$0.3 mJy \citep{gunawan93}.  This could 
result from the coliding winds of two massive stars.  

\begin{figure}
\epsscale{1.0}
\centering
\plotone{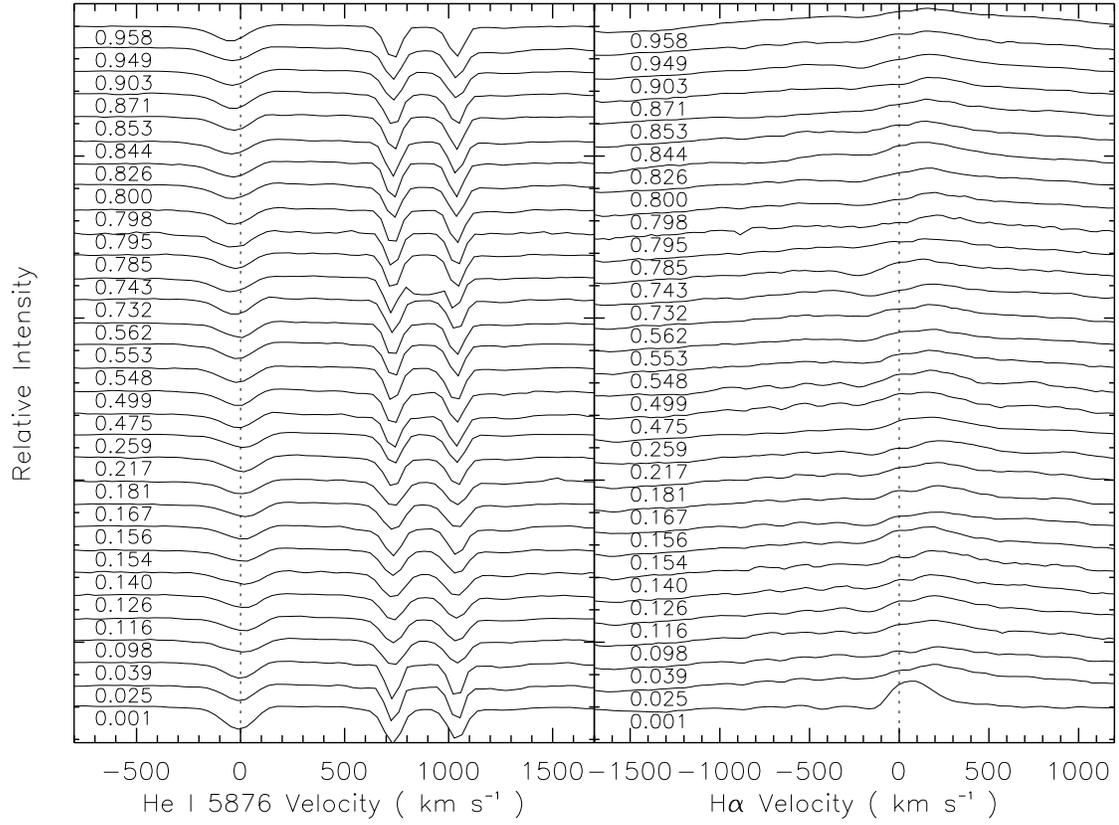}
\caption{A time sequence of spectra for MT734, ordered by phase, in the spectral region around
\ion{He}{1} $\lambda$5876 (left panel) and H$\alpha$ (right panel).  
\label{734sequence}}
\end{figure}

Figure~\ref{734sequence} shows the spectral time sequence for MT734,
ordered by phase, in the spectral region around \ion{He}{1}
$\lambda$5876 (left panel) and H$\alpha$ (right panel). The
\ion{He}{1} line shows a smooth progression of central wavelength
with orbital phase, while the H$\alpha$ line is broad, exhibits
emission, and has irregular line profiles, typical of supergiants.

\section{Discussion and Summary}

We have presented orbital solutions for five additional massive
binaries that are probable members of the Cygnus~OB2 Association,
bringing the total number of multiple systems therein to 25.  The vast
majority of these have complete orbital solutions, making this the
largest collection of massive binary statistics  in any single
cluster or association to date.  \citet{Kiminki2012b}  have used
these data, along with previously known systems documented in Papers
I--IV, to infer the underlying distribution of periods, mass ratios,
and eccentricities for Cyg~OB2 as a whole.    The analysis of
\citet{Kiminki2012b} suggests that the known list of 25 binaries out
of 114 Cyg~OB2 systems surveyed is reasonably complete at periods
less than about 30~days, but that we should expect to discover an
additional 10--15 binaries having longer periods as the Cygnus~OB2
Radial Velocity Survey continues. 

Noteworthy among the new list of massive binaries is the discovery of
a probable progenitor of a low-mass X-ray binary system (MT174).  
Also noteworthy is the long-period system MT070 with a period of 6.2
years.   Both of these low-amplitude systems ($K_1$$\simeq$9 \kms)
attest to the capabilities of the Survey to achieve orbital solutions
for  either low-inclination or low-$q$ systems.

\acknowledgements  We are grateful to Gregor Rauw who made us aware
of the possible binary nature of MT267.
 We thank the time allocation
committees of the Lick, Keck, WIYN, and WIRO observatories for
granting us observing time and making this project possible.  The
efforts of WIRO staff James Weger and Jerry Bucher made this science
possible. We acknowledge continued support from the National Science
Foundation through Research Experience for Undergraduates (REU)
program grant AST 03-53760, through grant AST 03-07778, and through
grant AST 09-08239, and the support of the Wyoming NASA Space Grant
Consortium through grant \#NNX10A095H.

\textit{Facilities:} \facility{WIRO ()}, \facility{WIYN ()}, \facility{Keck:I ()}

\thispagestyle{empty}

\clearpage

\begin{deluxetable}{lcrrrrr}
\centering
\tabletypesize{\tiny}
\tabletypesize{\scriptsize}
\tablewidth{0pc}
\tablecaption{Ephemeris for MT103 \label{103.tab}}
\tablehead{
\colhead{} & 
\colhead{} & 
\colhead{$V_{r1}$} &
\colhead{$O_1-C_1$} &
\colhead{$V_{r2}$} &
\colhead{$O_2-C_2$} \\ 
\colhead{Date (HJD-2,400,000)} &
\colhead{$\phi$} &
\colhead{(\kms)} &
\colhead{(\kms)} &
\colhead{(\kms)} &
\colhead{(\kms)} &
\colhead{Observatory}}
\tablecolumns{7}
\startdata
\cutinheadc{MT103}
 52146.75................... &     0.660    &    -12.0  (    5.0 )  &      1.5    &   -179.0  (   50.0 )  &   -128.6  &  WIYN  \\
 52162.75................... &     0.383    &   -102.9  (    8.0 )  &     -6.5    &     28.0  (   19.0 )  &    -37.4  &  WIYN  \\
 53989.75................... &     0.036    &     18.0  (    9.0 )  &      7.4    &   -186.0  (   33.0 )  &   -102.0  &  WIYN  \\
 54696.75................... &     0.031    &      7.0  (    6.0 )  &     -4.3    &   -117.0  (   23.0 )  &    -32.0  &  WIRO  \\
 54697.75................... &     0.075    &      6.0  (    4.0 )  &      1.2    &   -142.0  (   45.0 )  &    -66.1  &  WIRO  \\
 55779.75................... &     0.018    &      0.0  (    9.0 )  &    -12.8    &    -99.0  (   50.0 )  &    -11.9  &  WIRO  \\
 55791.50................... &     0.551    &    -55.0  (    7.4 )  &     27.5    &     41.0  (   13.0 )  &     -5.0  &  WIRO  \\
 55805.50................... &     0.186    &    -16.8  (   12.9 )  &      2.2    &    -55.0  (   14.0 )  &    -12.2  &  WIRO  \\
 55832.50................... &     0.403    &   -107.0  (    4.4 )  &     -1.7    &     71.1  (    9.0 )  &     -6.8  &  WIRO  \\
 55834.50................... &     0.495    &   -117.7  (    6.7 )  &     -2.0    &     80.5  (   12.0 )  &    -11.8  &  WIRO  \\
 55855.25................... &     0.441    &   -113.4  (    6.9 )  &      5.0    &     90.0  (   11.0 )  &     -6.1  &  WIRO  \\
 55857.50................... &     0.535    &    -99.0  (    5.1 )  &     -4.4    &     84.7  (   11.0 )  &     21.8  &  WIRO  \\
 55866.50................... &     0.941    &     32.5  (    7.6 )  &     12.6    &    -76.0  (   34.0 )  &     21.0  &  WIRO  \\
 55901.00................... &     0.511    &   -107.0  (    5.0 )  &      2.1    &    101.0  (   13.0 )  &     17.9  &  WIRO  \\
 55906.25................... &     0.744    &     16.0  (   11.5 )  &      4.3    &    -92.8  (   38.5 )  &     -7.3  &  WIRO  \\

\enddata 

\end{deluxetable}

\clearpage

\begin{deluxetable}{lcrrr}
\centering
\tabletypesize{\tiny}
\tabletypesize{\scriptsize}
\tablewidth{0pc}
\tablecaption{Ephemerides for MT070, MT174, MT267, \&\ MT734 \label{rest.tab}}
\tablehead{
\colhead{} & 
\colhead{} & 
\colhead{$V_{r1}$} &
\colhead{$O_1-C_1$} \\
\colhead{Date (HJD-2,400,000)} &
\colhead{$\phi$} &
\colhead{(\kms)} &
\colhead{(\kms)} &
\colhead{Observatory}}
\tablecolumns{5}
\startdata  
\cutinheadc{MT070}
 51467.00................... &     0.165    &    -19.7  (    2.1 )  &     -1.5  &  KECK   \\
 51806.00................... &     0.315    &     -8.7  (    1.8 )  &      2.7  &  KECK   \\
 52146.75................... &     0.465    &     -7.3  (    5.4 )  &      0.7  &  WIYN   \\
 52161.75................... &     0.472    &     -8.8  (    4.4 )  &     -0.9  &  WIYN   \\
 52162.75................... &     0.472    &    -10.0  (    4.5 )  &     -2.1  &  WIYN   \\
 53340.50................... &     0.994    &    -26.4  (    6.3 )  &     -2.4  &  WIYN   \\
 53989.75................... &     0.281    &    -18.0  (    9.9 )  &     -5.5  &  WIYN   \\
 54399.75................... &     0.462    &     -9.1  (    4.8 )  &     -1.1  &  WIRO   \\
 54401.75................... &     0.463    &    -16.1  (    3.3 )  &     -8.1  &  WIRO   \\
 54402.75................... &     0.464    &     -6.8  (    3.6 )  &      1.2  &  WIRO   \\
 54410.75................... &     0.467    &     -8.1  (    4.2 )  &     -0.2  &  WIRO   \\
 54696.75................... &     0.594    &     -5.2  (    2.9 )  &      1.7  &  WIRO   \\
 54697.75................... &     0.594    &     -7.7  (    3.3 )  &     -0.8  &  WIRO   \\
 54698.75................... &     0.595    &     -6.5  (    2.7 )  &      0.4  &  WIRO   \\
 54757.75................... &     0.621    &     -6.3  (    2.8 )  &      0.6  &  WIRO   \\
 55404.75................... &     0.907    &     -9.1  (    8.6 )  &      7.4  &  WIRO   \\
 55422.00................... &     0.915    &    -10.8  (    9.3 )  &      6.3  &  WIRO   \\
 55468.75................... &     0.935    &    -26.1  (   12.2 )  &     -7.0  &  WIRO   \\
 55491.50................... &     0.945    &    -22.0  (    3.4 )  &     -1.9  &  WIRO   \\
 55716.75................... &     0.045    &    -24.4  (    2.6 )  &      0.5  &  WIRO   \\
 55718.75................... &     0.046    &    -24.1  (    2.4 )  &      0.8  &  WIRO   \\
 55766.75................... &     0.067    &    -23.9  (    2.2 )  &      0.2  &  WIRO   \\
 55805.50................... &     0.084    &    -23.2  (    2.2 )  &      0.0  &  WIRO   \\
 55832.50................... &     0.096    &    -26.1  (    2.5 )  &     -3.6  &  WIRO   \\
 55834.50................... &     0.097    &    -22.2  (    2.1 )  &      0.3  &  WIRO   \\
 55847.50................... &     0.103    &    -22.0  (    2.1 )  &      0.1  &  WIRO   \\
 55855.25................... &     0.106    &    -22.7  (    2.6 )  &     -0.9  &  WIRO   \\
 55866.50................... &     0.111    &    -19.7  (    2.2 )  &      1.9  &  WIRO   \\
\cutinhead{MT174}							     
 55727.00................... &     0.974    &      7.3  (    4.3 )  &     -1.2  &  WIRO   \\
 55759.00................... &     0.029    &     16.7  (    1.9 )  &     -2.1  &  WIRO   \\
 55781.00................... &     0.879    &      6.0  (    2.9 )  &      3.0  &  WIRO   \\
 55791.50................... &     0.194    &     19.7  (    3.1 )  &      2.0  &  WIRO   \\
 55832.50................... &     0.233    &     15.0  (    2.6 )  &     -1.5  &  WIRO   \\
 55836.50................... &     0.115    &     20.9  (    2.5 )  &      0.7  &  WIRO   \\
 55855.50................... &     0.304    &     13.4  (    2.8 )  &     -1.2  &  WIRO   \\
 55857.50................... &     0.745    &      4.8  (    2.6 )  &     -0.6  &  WIRO   \\
 55876.25................... &     0.879    &      1.0  (    2.6 )  &     -2.1  &  WIRO   \\
 55906.25................... &     0.493    &     11.2  (    2.5 )  &      0.8  &  WIRO   \\
\cutinhead{MT267}							        
 55713.75................... &     0.950    &     19.2  (    4.2 )  &      4.9  &  WIRO   \\
 55714.75................... &     0.009    &      7.7  (    4.0 )  &     -0.9  &  WIRO   \\
 55715.75................... &     0.075    &     -7.1  (    4.2 )  &     -3.2  &  WIRO   \\
 55716.75................... &     0.138    &    -16.2  (    4.3 )  &     -0.1  &  WIRO   \\
 55717.75................... &     0.202    &    -19.1  (    4.3 )  &      6.2  &  WIRO   \\
 55718.75................... &     0.269    &    -25.2  (    4.3 )  &      5.5  &  WIRO   \\
 55726.75................... &     0.783    &      6.1  (    4.0 )  &      5.7  &  WIRO   \\
 55727.75................... &     0.850    &      6.1  (    3.2 )  &     -2.8  &  WIRO   \\
 55737.75................... &     0.493    &    -27.5  (    4.2 )  &      1.3  &  WIRO   \\
 55738.75................... &     0.558    &    -19.3  (    3.3 )  &      5.1  &  WIRO   \\
 55739.50................... &     0.601    &    -24.3  (   16.7 )  &     -3.5  &  WIRO   \\
 55740.50................... &     0.665    &    -12.7  (    3.7 )  &      1.6  &  WIRO   \\
 55741.50................... &     0.735    &    -11.1  (    3.3 )  &     -5.1  &  WIRO   \\
 55758.75................... &     0.846    &     13.3  (    3.0 )  &      5.0  &  WIRO   \\
 55765.75................... &     0.294    &    -35.8  (    4.2 )  &     -4.0  &  WIRO   \\
 55767.50................... &     0.406    &    -41.0  (    5.0 )  &     -8.7  &  WIRO   \\
 55768.50................... &     0.472    &    -30.6  (    4.8 )  &     -0.8  &  WIRO   \\
 55779.75................... &     0.192    &    -34.7  (    5.6 )  &    -10.6  &  WIRO   \\
 55779.75................... &     0.200    &    -18.3  (    4.8 )  &      6.8  &  WIRO   \\
 55780.75................... &     0.257    &    -35.8  (    5.9 )  &     -5.8  &  WIRO   \\
 55781.75................... &     0.321    &    -37.9  (    9.1 )  &     -5.4  &  WIRO   \\
 55790.50................... &     0.889    &      6.5  (    3.8 )  &     -6.1  &  WIRO   \\
 55834.50................... &     0.725    &    -10.1  (    4.0 )  &     -2.8  &  WIRO   \\
 55857.25................... &     0.199    &    -26.5  (    4.3 )  &     -1.6  &  WIRO   \\
 55866.25................... &     0.781    &      1.5  (    4.5 )  &      1.5  &  WIRO   \\
\cutinhead{MT734}							         
 54347.75................... &     0.025    &    -22.8  (   10.0 )  &     -1.1  &  WIRO   \\
 54348.75................... &     0.039    &    -17.4  (   10.0 )  &     -1.4  &  WIRO   \\
 54399.75................... &     0.743    &    -61.2  (   16.7 )  &    -16.3  &  WIRO   \\
 54402.75................... &     0.785    &    -42.2  (    8.3 )  &      6.1  &  WIRO   \\
 54403.50................... &     0.795    &    -54.9  (   10.0 )  &     -5.8  &  WIRO   \\
 54403.75................... &     0.798    &    -46.6  (   10.0 )  &      2.8  &  WIRO   \\
 54405.75................... &     0.826    &    -56.3  (   11.1 )  &     -4.5  &  WIRO   \\
 54642.75................... &     0.098    &     -8.8  (    5.3 )  &     -2.6  &  WIRO   \\
 54644.00................... &     0.116    &     -6.1  (    5.6 )  &     -0.1  &  WIRO   \\
 54644.75................... &     0.126    &     -6.5  (    5.3 )  &     -0.4  &  WIRO   \\
 54645.75................... &     0.140    &     -5.4  (    5.6 )  &      1.1  &  WIRO   \\
 54646.75................... &     0.154    &    -14.3  (    5.9 )  &     -7.3  &  WIRO   \\
 54647.75................... &     0.167    &     -0.5  (    5.3 )  &      7.3  &  WIRO   \\
 54648.75................... &     0.181    &    -14.7  (    5.3 )  &     -6.2  &  WIRO   \\
 54670.00................... &     0.475    &    -30.0  (    8.3 )  &     -3.3  &  WIRO   \\
 54671.75................... &     0.499    &    -33.6  (    7.7 )  &     -5.3  &  WIRO   \\
 54696.75................... &     0.844    &    -62.7  (    9.1 )  &     -9.4  &  WIRO   \\
 54747.75................... &     0.548    &    -33.5  (    8.3 )  &     -2.2  &  WIRO   \\
 54748.75................... &     0.562    &    -26.4  (    8.3 )  &      5.8  &  WIRO   \\
 55404.75................... &     0.619    &    -34.6  (   12.5 )  &      1.3  &  WIRO   \\
 55409.75................... &     0.688    &    -62.8  (   14.3 )  &    -22.1  &  WIRO   \\
 55421.75................... &     0.854    &    -46.7  (   14.3 )  &      7.4  &  WIRO   \\
 55422.75................... &     0.868    &    -50.1  (   14.3 )  &      5.1  &  WIRO   \\
 55423.75................... &     0.882    &    -46.1  (   14.3 )  &     10.0  &  WIRO   \\
 55461.75................... &     0.406    &    -22.4  (   20.0 )  &      0.2  &  WIRO   \\
 55466.75................... &     0.475    &    -24.8  (   20.0 )  &      2.0  &  WIRO   \\
 55498.75................... &     0.917    &    -44.3  (   20.0 )  &     13.1  &  WIRO   \\
 55715.00................... &     0.903    &    -57.9  (    5.0 )  &     -0.7  &  WIRO   \\
 55719.00................... &     0.958    &    -62.8  (    7.1 )  &     -9.9  &  WIRO   \\
 55737.75................... &     0.217    &    -10.2  (    6.2 )  &      0.5  &  WIRO   \\
 55740.75................... &     0.259    &     -7.5  (    5.9 )  &      5.8  &  WIRO   \\
 55780.00................... &     0.800    &    -54.4  (    6.7 )  &     -4.8  &  WIRO   \\
 55790.75................... &     0.949    &    -55.2  (    5.3 )  &     -0.3  &  WIRO   \\
 55794.50................... &     0.001    &    -20.7  (   12.5 )  &     14.1  &  WIRO   \\
 55805.75................... &     0.156    &     -9.2  (    5.9 )  &     -2.0  &  WIRO   \\
 55834.50................... &     0.553    &    -32.5  (    8.3 )  &     -0.9  &  WIRO   \\
 55847.50................... &     0.732    &    -44.0  (    8.3 )  &      0.1  &  WIRO   \\
 55856.25................... &     0.853    &    -51.9  (    8.3 )  &      2.2  &  WIRO   \\
 55857.50................... &     0.871    &    -49.9  (    7.1 )  &      5.5  &  WIRO   \\

\enddata 

\end{deluxetable}

\clearpage
 
\begin{deluxetable}{lrrrrrr}
\centering
\tabletypesize{\scriptsize}
\tablewidth{0pt}
\tablecaption{Orbital Elements \label{solutions.tab}}
\tablehead{
\colhead{Element} &
\colhead{MT070} &
\colhead{MT103} &
\colhead{MT174} &
\colhead{MT267} & 
\colhead{MT734} }  
\startdata            
$P$ (Days)                & 2259$\pm$46       & 22.104$\pm$0.002   & 4.536$\pm$0.020   & 15.511$\pm$0.056  & 72.43$\pm$0.07 \\
$e$                       & 0.34$\pm$0.11     & 0.32$\pm$0.05      & 0.53$\pm$0.13     & 0.21$\pm$0.07     & 0.50$\pm$0.06    \\
$\omega$ (deg)            & 157.9$\pm$26      & 26$\pm$12          & 282.0$\pm$20      & 37$\pm$19         & 262.6$\pm$8      \\
$\gamma$ (\kms)           & $-$13$\pm$1         & $-$28$\pm$3          & $-$6$\pm$2          & $-$13$\pm$1         & $-$31 $\pm$1       \\
$T_0$ (HJD-2,400,000)     & 53355$\pm$134     & 55757.2$\pm$0.5    & 55749.8$\pm$0.1   & 55745.7$\pm$0.8   & 55722$\pm$1   \\
$K_{1}$ (\kms)            & 9$\pm$1           & 71$\pm$4           & 9$\pm$1           & 24$\pm$2          & 26$\pm$1         \\
$K_{2}$ (\kms)            & \nodata           & 100$\pm$8          & \nodata           & \nodata	   & \nodata          \\
S.~C.$_1$                 & O9III              & B1V                & B2III             & O7.5III--I        & O5I(f)           \\
S.~C.$_2$                 & B:                 & B2V                & \nodata          & B -- O	           & early B -- O      \\
$q$                       & 0.22--1           & 0.71$\pm$0.10      & 0.03--0.9         & 0.09--1           & 0.15--1          \\
$i$  (degrees)            & 90--18            & $\simeq$50         & 87--3             & 81--8             & 90--13           \\
$a$ (AU)                  & 9.3--10.9         & 0.45$\pm$0.02      & 0.12--0.15        & 0.39--0.48        & 1.3--1.5          \\
$\chi_{red}^2$            & 0.68              & 2.41               & 0.71              & 1.59	           & 0.54     \\

\enddata

\end{deluxetable} 
 
\clearpage

\begin{deluxetable}{lcccccl}
\centering
\tabletypesize{\scriptsize}
\setlength{\tabcolsep}{0.06in}
\tablewidth{0pc}
\tablecaption{OB Binaries in Cyg OB2 \label{Binaries6}}
\tablehead{
\colhead{} & 
\colhead{} & 
\colhead{} & 
\colhead{P} &
\colhead{} &
\colhead{} &
\colhead{} \\
\colhead{Star} &
\colhead{Type} &
\colhead{S.C.} &
\colhead{(days)} &
\colhead{e} &
\colhead{q} &
\colhead{Ref.}}
\startdata  
MT059      & SB1     & O8V \&\ B                          & 4.8527$\pm$0.0002    & 0.11$\pm$0.04   & 0.22--0.67  		& 1 \\
MT070      & SB1     & O9III \& B                         & 6.19~yr              & 0.34$\pm$0.11   & 0.25--1                    & 0 \\
MT103      & SB2     & B1V $+$ B2?                        & 22.104$\pm$0.002     & 0.32$\pm$0.05   & 0.71$\pm$0.10              & 0 \\
MT145      & SB1     & O9III \&\ mid B                    & 25.140$\pm$0.008     & 0.291$\pm$0.009 & 0.26--0.63  		& 2 \\
MT174      & SB1     & B2III  \& ??                       & 4.536$\pm$0.020      & 0.53$\pm$0.13   & 0.03--0.9                  & 0 \\
MT252      & SB2     & B2III \&\ B1V                      & 18--19               & \nodata         & 0.8$\pm$0.2                & 1 \\
MT258      & SB1     & O8V \&\ B                          & 14.660$\pm$0.002     & 0.03$\pm$0.05   & 0.18--0.89  		& 1 \\
MT267      & SB1     & O7.5III-I \& O/B                   & 15.511$\pm$0.056     & 0.21$\pm$0.07   & 0.09--1                    & 0 \\
MT311      & SB2     & B2V \&\ B3V                        & 5.7521$\pm$0.0002    & 0.02$\pm$0.01   & 0.8$\pm$0.1                & 3  \\
MT372      & SB1/EA: & B0V \&\ B2:V                       & 2.228 (fixed)        & 0.0 (fixed)     & $\sim$0.6                 	& 2,4 \\
MT421      & SB1:/EA & O9V \&\ B9V--A0V                   & 4.161   	         & \nodata         & $\sim$0.16--0.19          	& 5 \\
MT429      & SB1/EA  & B0V \&\ G0III:                     & 2.9788 (fixed) 	 & 0.38$\pm$0.08   & 0.057:                     & 3,5 \\
MT605      & SB2     & B1V  \&\ B1:                       & $\sim$4--5           & \nodata         & 0.9$\pm$0.1                & 3 \\
MT696  & SB2/EW/KE   & O9.5V \&\ B0V                      & 1.4694$\pm$0.0002	 & 0.081$\pm$0.005 & 0.94$\pm$0.03              & 3,6 \\
MT720      & SB2     & B0--B1 \&\ B1--B2                  & 4.0677$\pm$0.0003    & 0.35$\pm$0.02   & 0.80$\pm$0.08              & 1,3 \\
MT734      & SB1     & O5I \& O/early B                   & 72.43$\pm$0.07       & 0.50$\pm$0.06   & 0.15 --1                   & 0\\
MT771      & SB2     & O7V \&\ O9V                        & 2.82105$\pm$0.00003  & 0.547$\pm$0.004 & 0.95$\pm$0.09	        & 1,3 \\
Schulte 3  & SB2/EA: & O6IV: \&\ O9III                    & 4.7464$\pm$0.0002    & 0.070$\pm$0.009 & 0.44$\pm$0.08              & 1,7 \\
Schulte 5  & SB2/EB  & O7Ianfp \&\ Ofpe/WN9               & 6.6 (fixed)          & 0.0 (fixed)     & 0.28$\pm$0.02            	& 8,9,10,11, \\
           &         & (\&\ B0V:)                         &                      &                 &                            & 12,13,14 \\
Schulte 8a & SB2     & O5.5I \&\ O6:                      & 21.908 (fixed)       & 0.24$\pm$0.04   & 0.86$\pm$0.04            	& 15,16 \\
Schulte 9  & SB2     & O5: \&\ O6--7:                     & 2.355~yr             & 0.708$\pm$0.027 & 0.9$\pm$0.1                & 17,18      \\
Schulte 73 & SB2     & O8III \&\ O8III                    & 17.28$\pm$0.03       & 0.169$\pm$0.009 & 0.99$\pm$0.02              & 2      \\
A36        & SB2/EA  & B0Ib \&\ B0III                     & 4.674$\pm$0.004      & 0.10$\pm$0.01   & 0.70$\pm$0.06              & 2,19,20   \\
A45        & SB2     & B0.5V \&\ B2V:--B3V:               & 2.884$\pm$0.001      & 0.273$\pm$0.002 & 0.46$\pm$0.02              & 2,20      \\
B17        & SB2/EB: & O7: \&\ O9:                        & 4.0217$\pm$0.0004    & 0 (fixed)       & 0.75 (fixed)             	& 21,22     \\
\enddata

\tablecomments{Photometric types EW/KE, EA, and EB stand for eclipsing
system of the W UMa type (ellipsoidal; $P<1$ day), Algol type (near
spherical), and $\beta$ Lyr type (ellipsoidal; $P>1$ day)
respectively. The mass ratio for MT421 is calculated using the O star
masses of \citet{martins05} and interpolated AB masses of
\citet{drilling}. }

\tablerefs{
(0) This work;
(1) Paper~II; 
(2) Paper~III;
(3) Paper~IV;
(4) \citet{Wozniak04};
(5) \citet{PJ98}; 
(6) \citet{Rios04};
(7) \citet{Kiminki2012b}; 
(8) \citet{Wilson48}; 
(9) \citet{Wilson51}; 
(10) \citet{Mics53};
(11) \citet{Wal73}; 
(12) \citet{Contreras97}; 
(13) \citet{Rauw99}; 
(14) \citet{Hall74};
(15) \citet{Romano69}; 
(16) \citet{Debeck04};
(17) \citet{Naze08};
(18) \citet{Naze10};
(19) \citet{NSVSa};
(20) \citet{Hanson03};
(21) \citet{Stroud10};
(22) \citet{NSVSb}}

\end{deluxetable}

\clearpage

\clearpage

\clearpage

\clearpage

\clearpage

\clearpage

\clearpage

\clearpage

\clearpage

\clearpage

\clearpage

\clearpage

\clearpage

\clearpage

\clearpage


\begin{thebibliography}{}
\bibitem[Bohlin et a.(1978)]{bohlin78} Bohlin, R.~C., Savage, B.~D., \& Drake, J.~F. 1978, ApJ, 224, 132
\bibitem[Burrows et al.(1995)]{burrows95} Burrows, A., Hayes, J., \& Fryxell, B.~A.\ 1995, \apj, 450, 830 
\bibitem[Blaauw(1961)]{blaauw} Blaauw, A. 1961, BAN, 15, 265
\bibitem[Caballero-Nieves et al.(2011)]{2011BSRSL..80..639C} Caballero-Nieves, S.~M., Gies, D.~R., Roberts, L.~C., \& Turner, N.~H.\ 2011, Bulletin de la Societe Royale des Sciences de Liege, 80, 639
\bibitem[Comer{\'o}n et al.(2002)]{comeron02} Comer{\'o}n, F., Pasquali, A., Rodighiero, G., et al.\ 2002, \aap, 389, 874 
\bibitem[Contreras et al.(1997)]{Contreras97} Contreras, M. E., Rodriguez, L. F., Tapia, M., Cardini, D., Emanuele, A.,  Badiali, M., \&\ Persi, P. 1997, ApJ, 488, 153
\bibitem[De Becker et al.(2004)De Becker, Rauw, and Manfroid]{Debeck04} De Becker, M., Rauw, G., \&\ Manfroid, J. 2004, A\&A, 424, L39
\bibitem[Drilling \& Landolt(2000)]{drilling} Drilling, J. S., \& Landolt, A. U. 2000, in Astrophysical Quantities, ed. A. N. Cox (4th ed.; New York; Springer), 381
\bibitem[Eldridge et al.(2008)]{eldridge2008} Eldridge, J.~J., Izzard, R.~G., \& Tout, C.~A.\ 2008, \mnras, 384, 1109
\bibitem[Garmany et al.(1980)]{garmany80} Garmany, C.~D., Conti,  P.~S., \& Massey, P.\ 1980, \apj, 242, 1063 
\bibitem[Setia Gunawan et al.(2003)]{gunawan93} Setia Gunawan, D.~Y.~A., de Bruyn, A.~G., van der Hucht, K.~A., 
 \& Williams, P.~M.\ 2003, \apjs, 149, 123 
\bibitem[Gies \&\ Bolton(1986)]{giesbolton} Gies, D. R., \&\ Bolton, C. T. 1986, ApJS, 61, 419
\bibitem[Gray \&\ Corbally(2009)]{Gray2009} Gray, R. O., \&\ Corbally, C. J. 2009, in Stellar Spectral Classification, ed. Spergel, D. N. (1st ed.; New Jersey, Princeton University Press), 115
\bibitem[Hall(1974)]{Hall74} Hall, D. S. 1974, AcA, 24, 69
\bibitem[Hanson(2003)]{Hanson03} Hanson, M. M. 2003, ApJ, 597, 957
\bibitem[Henderson et al.(2011)]{Henderson2011} Henderson, C. B., Stanek, K. Z., \&\ Prieto, J. L. 2011, ApJ, 194, 27
\bibitem[Herrero et al.(2002)]{herrero2002} Herrero, A., Puls, J., \& Najarro, F.\ 2002, \aap, 396, 949 
\bibitem[Izzard et al.(2004)]{izzard2004} Izzard, R.G., Ramirez-Ruiz, E, Tout, C.A. 2004, MNRAS, 348, 1215
\bibitem[Jacoby et al.(1984)]{jacoby84} Jacoby, G.~H. \& Hunter, D.~A. 1984, \apjs, 56, 257 
\bibitem[Jenniskens \& D{\'e}sert(1994)]{jenniskens} Jenniskens, P., \& D{\'e}sert, F.~X. 1994, Astronomy \& Astrophysics Suppliment Series, 106, 39
\bibitem[Kiminki et al.(2007)]{Kiminki07} Kiminki, D. C., et al. 2007, ApJ, 664, 1120 (Paper I)
\bibitem[Kiminki et al.(2008)]{Kiminki08} Kiminki, D. C., McSwain, M. V., \&\ Kobulnicky, H. A. 2008, ApJ, 679, 1478 (Paper II)
\bibitem[Kiminki et al.(2009)]{Kiminki09} Kiminki, D. C., Kobulnicky, H. A., Gilbert, I., Bird, S., Chunev, G. 2009, AJ, 137, 4608 (Paper III)
\bibitem[Kiminki et al.(2012)]{Kiminki2012a} Kiminki, D.~C., Kobulnicky, H.~A., Ewing, I., et al.\ 2012, \apj, 747, 41 (Paper IV)
\bibitem[Kiminki \&\ Kobulnicky(2012)]{Kiminki2012b} Kiminki, D. C., \&\ Kobulnicky, H. A. 2012, ApJ, 751, 4 (Paper V)
\bibitem[Kobulnicky \& Fryer(2007)]{kf07} Kobulnicky, H.~A., \& Fryer, C.~L.\ 2007, \apj, 670, 747 
\bibitem[Krumholz et al.(2010)]{Krumholzetal2010} Krumholz, M.~R., Cunningham, A.~J., Klein, R.~I., \& McKee, C.~F.\ 2010, \apj, 713, 1120 
\bibitem[Krumholz \&\ Thompson(2007)]{Krumh} Krumholz, M. R., \&\ Thompson, T. A. 2007, ApJ, 661, 1034
\bibitem[Hubeny  \& Lanz(1995)]{hubeny} Hubeny, I., \& Lanz, T.\ 1995, \apj, 439, 875 
\bibitem[Lanz \&\ Hubeny(2003)]{lanz} Lanz, T., \&\ Hubeny, I. 2003, ApJS, 146, 417
\bibitem[Maeder \& Meynet(2000)]{meynet2000} Maeder, A., \& Meynet, G.\ 2000, \araa, 38, 143
\bibitem[Markwardt(2009)]{mpfit} Markwardt, C.~B.\ 2009, Astronomical Data Analysis Software and Systems XVIII, 411, 251 
\bibitem[Martins et al.(2005)]{martins05} Martins, F., Schaerer, D., \& Hillier, D.~J. 2005, \aap, 436, 1049
\bibitem[Massey \&\ Thompson(1991)]{MT91} Massey, P., \&\  Thompson, A. B. 1991, AJ, 101, 1408
\bibitem[Miczaika(1953)]{Mics53} Miczaika, G. R. 1953, PASP, 65, 141
\bibitem[Naz\'{e} et al.(2008)]{Naze08} Naz\'{e}, Y., De Becker, M., Rauw, G., \&\ Barbieri, C. 2008, A\&A, 483, 543 
\bibitem[Naz\'{e} et al.(2010)]{Naze10} Naz\'{e} et al. 2010, ApJ, 719, 634
\bibitem[Negueruela et al.(2008)]{Negueruela08} Negueruela, I., Marco, A., Herrero, A., \& Clark, J.~S.\ 2008, \aap, 487, 575 
\bibitem[Otero(2008a)]{NSVSa} Otero, S. 2008a, Open European Journal on Variable Stars, 83, 1
\bibitem[Otero(2008b)]{NSVSb} Otero, S. 2008b, Open European Journal on Variable Stars, 91, 1
\bibitem[Pigulski \&\ Kolaczkowski(1998)Pigulski \&\ Kolaczkowski]{PJ98} Pigulski, A., \&\ Kolaczkowski, Z. 1998, MNRAS, 298, 753
\bibitem[Poppi et al.(2010)]{poppi} Poppi, S., Scappini, F., Cecchi-Pestellini, C., \& Maccaferri, G.\ 2010, \mnras, 407, 1255 
\bibitem[Rauw et al.(1999)]{Rauw99} Rauw, G., Vreux, J. M., \&\ Bohannan, B. 1999, ApJ 517, 416
\bibitem[Rauw(2011)]{rauw2011} Rauw, G.\ 2011, \aap, 536, A31
\bibitem[Rios \&\ DeGioia-Eastwood(2004)]{Rios04} Rios, L. Y., \&\ DeGioia-Eastwood, K. 2004, BAAS, 205, No. 09.05
\bibitem[Roberts et al.(1987)]{Roberts87}
Roberts, D. H., Leh\'{a}r, J., \& Dreher, J. W. 1987, AJ, 93, 968 
\bibitem[Romano(1969)]{Romano69} Romano, G. 1969, MmSAI, 40, 375
\bibitem[Salpeter(1955)]{salpeter55} Salpeter, E. E. 1955, ApJ, 121, 161
\bibitem[Sana \& Evans(2011)]{sanaevans2011} Sana, H., \& Evans,  C.~J.\ 2011, IAU Symposium, 272, 474
\bibitem[Scappini et al.(2007)]{scappini} Scappini, F., Cecchi-Pestellini, C., Casu, S., \& Olberg, M.\ 2007, \aap, 466, 243 
\bibitem[Schulte(1958)]{Schulte58} Schulte, D.~H. 1958, AJ, 128, 41
\bibitem[Smith et al.(2009)]{smithetal2009} Smith, R.~J., Longmore, S., \& Bonnell, I.\ 2009, \mnras, 400, 1775 
\bibitem[Solomon et al.(1987)]{solomon87} Solomon, P.~M., Rivolo, A.~R., Barrett, J., \& Yahil, A.\ 1987, \apj, 319, 730 
\bibitem[Sota et al.(2011)]{Sota2011} Sota, A., Ma\'{i}z Apell\'{a}niz, J., Walborn, N. R., Alfaro, E. J., Barb\'{a}, R. H., Morrell, N. I., Gamen, R. C., \&\ Arias, J. I. 2011, 193, 24
\bibitem[Stroud et al.(2010)]{Stroud10} Stroud, V. E., Clark, J.S., Negueruela, I. , Roche, P.,  \&\ Norton, A.J. 2009, A\&A, 511, 84
\bibitem[van den Heuvel(1983)]{vanden83} van den Heuvel,  E.~P.~J.\ 1983, Accretion-Driven Stellar X-ray Sources, 303
\bibitem[Walborn(1973)]{Wal73} Walborn, N. R. 1973, ApJ, 180, L35
\bibitem[Walborn(1980)]{Walborn80} Walborn, N. R. 1980, ApJS, 44, 535
\bibitem[Walborn(2009)]{Walborn2009} Walborn, N. R. 2009, in Stellar Spectral Classification, ed. Spergel, D. N. (1st ed.; New Jersey, Princeton University Press), 66
\bibitem[Walborn \&\ Fitzpatrick(1990)]{Walborn90} Walborn, N. R., \&\ Fitzpatrick, E. L. 1990, PASP, 102, 379
\bibitem[Wilson(1948)]{Wilson48} Wilson, O. C. 1948, PASP, 60, 385
\bibitem[Wilson \&\ Abt(1951)]{Wilson51} Wilson, O. C., \&\ Abt, A. 1951, ApJ, 144, 477
\bibitem[Woosley  \& Bloom(2006)]{woosley2006} 
 Woosley, S.~E., \& Bloom, J.~S.\ 2006, \araa, 44, 507 
\bibitem[Woosley et al.(1993)]{woosleylangerweaver} Woosley, S.~E., Langer, 
N., \& Weaver, T.~A.\ 1993, \apj, 411, 823 
\bibitem[Wozniak et al.(2004)]{Wozniak04} Wozniak, P. R., et al. 2004, AJ, 127, 2436,  Northern Sky Variability Survey: Public Data Release 

\end{thebibliography}
\end{document}